\documentclass[tradiabstract]{aa} 
\usepackage{graphicx}
\usepackage{supertabular}
\usepackage{txfonts}
\begin{document}
\title{Binary planetary nebulae nuclei towards the Galactic bulge\thanks{Based on observations made with Gemini South under programs GS-2008B-Q-65 and GS-2009A-Q-35.}}

   \subtitle{II. A penchant for bipolarity and low-ionisation structures}

   \author{B. Miszalski
           \inst{1,2}
           \and
           A. Acker
           \inst{1}
           \and
           Q. A. Parker
           \inst{2,3}
           \and
           A. F. J. Moffat
           \inst{4}
          }

\institute{
Observatoire Astronomique, Universit\'e de Strasbourg, 67000, Strasbourg, France\\
\email{[brent;acker]@newb6.u-strasbg.fr}
\and
Department of Physics, Macquarie University, Sydney, NSW 2109, Australia\\ 
             \email{[brent;qap]@ics.mq.edu.au}
         \and
         Anglo-Australian Observatory, Epping, NSW 1710, Australia\\
         \email{qap@ics.mq.edu.au}
         \and
         D\'ept. de physique, Univ. de Montr\'eal C.P. 6128, Succ. Centre-Ville, Montr\'eal, QC H3C 3J7, and Centre de recherche\\ en astrophysique du Qu\'ebec, Canada\\
         \email{moffat@astro.umontreal.ca}
         }
   \date{Received -; accepted -}
   \abstract{
   Considerable effort has been applied towards understanding the precise shaping mechanisms responsible for the diverse range of morphologies exhibited by planetary nebulae (PNe). 
   A binary companion is increasingly gaining support as a dominant shaping mechanism, however morphological studies of the few PNe that we know for certain were shaped by binary evolution are scarce or biased.  
  Newly discovered binary central stars (CSPN) from the OGLE-III photometric variability survey have significantly increased the sample of post common-envelope (CE) nebulae available for morphological analysis. 
   We present Gemini South narrow-band images for most of the new sample to complement existing data in a qualitative morphological study of 30 post-CE nebulae. Nearly 30\% of nebulae have canonical bipolar morphologies, however this rises to 60\% once inclination effects are incorporated with the aid of geometric models. This is the strongest observational evidence yet linking CE evolution to bipolar morphologies. A higher than average proportion of the sample shows low-ionisation knots, filaments or jets suggestive of a binary origin. These features are also common around emission-line nuclei which may be explained by speculative binary formation scenarios for H-deficient CSPN. 
   }

   \keywords{ISM: planetary nebulae: general - binaries: close - stars: Wolf-Rayet}
   \maketitle
\section{Introduction}
 
A single shaping mechanism behind the extraordinary morphological diversity of planetary nebulae (PNe) remains elusive even after more than 20 years of discussion (Balick \& Frank 2002). Highly axisymmetric, non-spherical forms in particular are very difficult to explain without a binary engine that is capable of supplying sufficient angular momentum to heavily shape the nebula (Nordhaus, Blackman \& Frank 2007). The \emph{binary hypothesis} encapsulated by De Marco (2009) simply states that PNe derive from binary progenitors \emph{more easily} than from single progenitors.
Gathering the \emph{direct evidence} of binary central stars (CSPN) required to test the binary hypothesis has however been a slow process with only around a dozen known until recently (De Marco, Hillwig \& Smith 2008) compared to $\sim$2600 PNe known in the Galaxy (Parker et al. 2006; Miszalski et al. 2008a).

Such a small sample has considerably hampered attempts to reveal morphological trends amongst the nebulae surrounding binary CSPN. We know their nebulae must have been ejected during the common-envelope (CE) phase (Iben \& Livio 1993) and attempts to model this process have suggested the geometry of the mass loss is subject to a density contrast between the equatorial and polar directions (Livio \& Soker 1988; Bond \& Livio 1990; Sandquist et al. 1998). A high density contrast is thought to produce bipolar nebulae and a mild one elliptical nebulae (Livio \& Soker 1988; Bond \& Livio 1990). The complexity of these models limits their predictive power when it comes to morphological studies of observed nebulae, however there are some indications that the density contrast could be as high as 5:1 (Sandquist et al. 1998). This remains to be proven with more advanced simulations, but if such a density contrast is accurate, then we may suspect to find post-CE nebulae dominated by bipolar morphologies.

It may therefore be surprising that Bond \& Livio (1990) did not find a large proportion of bipolar nebulae amongst $\sim$12 post-CE nebulae wherein only NGC 2346 satisfied the strict morphological criteria of canonical bipolar nebulae (Corradi \& Schwarz 1995). With such a small sample it is difficult to determine whether the theoretical expectations or the morphological classifications are responsible. We suspect projection effects are primarily responsible for the \emph{apparently} low bipolar fraction amongst post-CE PNe. Few morphological studies have attempted to account for the complex 3D structure of each nebula when projected onto the 2D plane of the sky (e.g. Manchado et al. 2004). The appearance of each nebula is therefore strongly dependent on its inclination angle $i$ with respect to the plane of the sky (e.g. Frank et al. 1993) and recovering the instrinsic 3D morphology requires detailed kinematic modelling. Consistent application of these techniques to post-CE nebulae has only just commenced with Mitchell et al. (2007) revealing an intrinsic bipolar morphology to A 63. 
Mitchell et al. (2007) demonstrated effectively the utility of knowing the orbital inclination of the binary CSPN in building a comprehensive model of the nebula in conjunction with kinematic data. 

Even without kinematic information there is still room for \emph{qualitative} classification within the framework of geometric bipolar models for post-CE nebulae. Pollacco \& Bell (1997) constructed such a model and made favourable comparisons with A 41, A 46, A 63 and A 65 suggesting they are bipolar nebulae. Similarly, De Marco (2009) considered inclination effects without the assistance of models to estimate as much as 70\% of post-CE nebulae could indeed be bipolar. This estimate is however disputable in that it was heavily biased by a small sample and in some cases the highest quality images available in the literature were not used (e.g. A 65). A detailed reappraisal of a large sample of post-CE nebulae is therefore required to substantiate the suspected high bipolar fractions from recent observations and to compare them with current theoretical expectations.  

To conduct this reappraisal we combine the small sample of prevously known post-CE nebulae (De Marco et al. 2008) with those found by Miszalski et al. (2008b; 2009, hereafter `Paper I'). In Paper I we cross-referenced previously known PNe and those from the Macquarie/AAO/Strasbourg H$\alpha$ (MASH) PNe catalogues (Parker et al. 2006; Miszalski et al. 2008a) with high quality time-series photometry from the OGLE-III survey towards the Galactic bulge (Udalski 2009). Our comprehensive approach yielded a more representative sample of 21 new periodic binary CSPN ideally suited to a less-biased morphological study. 

This paper is structured as follows. Section \ref{sec:obs} discusses our sample selection and available observations. Geometric models and a classification scheme are introduced in Sec. \ref{sec:scheme}. These are used to guide morphological classification of the sample in Sec. \ref{sec:morph} where individual PNe are discussed in detail. Section \ref{sec:penchant} describes the frequency of the morphological types, while sections  \ref{sec:originlis} and \ref{sec:enuclei} discuss low-ionisation structures. We conclude in Sec. \ref{sec:conclusion}. 

\section{Observations}
\label{sec:obs}
\subsection{Sample selection}

We are solely concerned with the spatially resolved nebulae surrounding periodic close binary central stars. Spectroscopic confirmation of each central star is required to construct a clean sample, however those without confirmation may also be accommodated depending upon the quality of their candidate CSPN. 
Central stars tabulated by De Marco (2009) that appear to be visual binaries, have composite spectra or main-sequence near-infrared colours are omitted from our sample since their periods are unknown. Visual binaries are not directly comparable to our sample because their influence on nebular shaping is expected to be less dramatic than in close binaries (Soker 1997), whereas unconfirmed composite or cool central stars without periods may just be superpositions (M\'endez 1989). 

Table \ref{tab:morph} lists 15 nebulae from De Marco et al. (2008) and 18 nebulae from Paper I that meet our selection criteria. These samples will be referred to as the known sample and the OGLE sample, respectively.
Column three gives the orbital period, column four the major and minor axis diameters, and column five indicates whether stellar eclipses occur in the lightcurve. Other columns are reserved for \emph{Spitzer} IRAC photometry (Sec. \ref{sec:mir}) or morphological classification purposes (Sec. \ref{sec:models}). 

The binary status of 14 members from the known sample is well established with plenty of observational evidence (De Marco et al. 2008). Only periodic variability is reported for HaTr 4 (Bond \& Livio 1990), but we include it in our sample based on the well-centred position of the CSPN. 
We exclude SuWt 2 from De Marco et al. (2008) because the inclination of the bipolar nebula deduced from the equatorial ring is inconsistent with the double A-type eclipsing central star and the true central star may yet turn out to be one of two other candidates revealed in high quality imaging (Smith et al. 2007). Also omitted is Hb 12 whose period is purportedly irregular (Hillwig, priv. communication to De Marco 2009), as well as the group of A 35-type binaries LoTr 1, LoTr 5 and A 35 since the PN status of A 35 is disputed (Frew 2008; De Marco 2009). 

We conducted spectroscopic observations of 14 binary CSPN candidates from Paper I under Gemini programs GS-2008B-Q-65 and GS-2009A-Q-35. The Gemini Multi-Object Spectrograph (GMOS, Hook et al. 2004) was used in its longslit configuration mostly with the B1200 grating to target the primary component over the wavelength range 4085--5550 \AA. Future papers will discuss the spectra in detail, however we can discuss here those either not confirmed, awaiting confirmation or having some other significance. Most candidates have been confirmed after identification of features belonging to the primary (HI and HeII absorption) or the irradiation of the secondary (CIII and NIII stellar emission, e.g. Pollacco \& Bell 1993, 1994). 
For example, features of both components were detected in K 6-34, leading us to revise downward its period to 0.20 days. No features of the primary were visible for PHR 1744$-$3355 and PHR 1801$-$2718, each of which shows G-type features shifted by at least 50 km/s from the nebula. 
Although we cannot rule out binary motion as the cause of the radial velocity difference, both associations seem unlikely at this stage and so we exclude them from our sample. A composite spectrum is however detected in PHR 1804$-$2913, but radial velocity monitoring will be required to confirm whether its unusual 6.7 day period is indeed an orbital period.
Further additions to our sample from Paper I include H 1-33, Bl 3-15 and PHR 1804$-$2645 that have well-located binary CSPN candidates that were too faint or too reddened to show any stellar features in their AAOmega spectra (described in Paper I). Another member of this group would be JaSt 66, but lacking any resolved imagery we have omitted it from our sample. The remainder of our sample includes PHR 1759$-$2915, MPA 1759$-$3007 and PHR 1801$-$2947 whose binary CSPN candidates are yet to be confirmed, however these will be treated separately from the other objects given their higher uncertainty.

\begin{table*}
   \centering
   \caption{PNe containing close binary central stars whose nebular morphologies are assessed in this work. 
   }
   \label{tab:morph}
   \begin{tabular}{llrrlrlllll}
      \hline\hline
PN G & Name  & Period  & Diameter & Eclipsing   &$i$ (lit.)&Model/$i$& Morphology & Traits &         IRAC & Ref.\\
&       & (days)     &     (arcsec)       & &(deg)&&     &      &   1\,\, 2\, 3\, 4 &        \\
\hline
005.0$+$03.0 & Pe 1-9          &0.14& 12 $\times$ 14& Y  &$\sim$90&-& MS &  L                        & N U N U &b\\
355.3$-$03.2 & PPA 1747$-$3435 &0.22& 12 & N  &-&-&  Irr & -                      & U U U U&a \\
355.7$-$03.0 & H 1-33          &1.13& 5 $\times$ 12 & N  &-&A/45& CB & G                        & U Y U \textbf{Y}  &a \\
354.5$-$03.9 & Sab 41          &0.30& 95 $\times$ 194 & N  &-&A/40& LB & L,J,R                        & U U U U        & b \\
000.6$-$01.3 & Bl 3-15         &0.27& 4 $\times$ 11 & N &- & A&  LB & G,J              & Y Y \textbf{Y} \textbf{Y} & a \\
359.1$-$02.3 & M 3-16          &0.57& 9 $\times$ 13& Y  &$\sim$90&-& Irr &  L,J                       & Y \textbf{Y} Y \textbf{Y} & b\\
357.6$-$03.3 & H 2-29          &0.24& 11 $\times$ 14 & Y  &$\sim$90&A/90& CB &  L,R                        & U U U U      &b \\
000.2$-$01.9 & M 2-19          &0.67& 7 $\times$ 19 & N  &-&A/45& CB &  G,R                        & Y Y Y \textbf{Y}& b \\
358.7$-$03.0 & K 6-34          &0.20& 14& N  &-&-&  MS/LB & G,L,J?,R              & N N N Y &b \\
357.0$-$04.4 & PHR 1756$-$3342 &0.26& 22 & N  &-&B/30&  PB & -                        & U U U U &b \\
001.8$-$02.0 & PHR 1757$-$2824 &0.80& 4 $\times$ 18 & Y  &$\sim$90&A/90&  CB & -                        & N N N N  &b \\
001.2$-$02.6 & PHR 1759$-$2915 &1.10& 25 $\times$ 30 & Y  &$\sim$90&A/90&  CB & -                        & N N N N &b \\
000.5$-$03.1a& MPA 1759$-$3007 &0.50& 17 $\times$ 42 & N  &-&A/55& LB & -                        & U N U N  &b \\
001.9$-$02.5 & PPA 1759$-$2834 &0.31& 17 $\times$ 35 & N  &-&B/45&  LB & I                        & N N N N &b \\
357.1$-$05.3 & BMP 1800$-$3408 &0.14& 30 $\times$ 55 & Y  &$\sim$90&-&  Irr & I                       & U U U U &b \\
000.9$-$03.3 & PHR 1801$-$2947 &0.32& 34 & N  &-&B/30&  PB & -                       & N N N N  &b \\
001.8$-$03.7 & PHR 1804$-$2913 &6.66& 3.5, 8 & N  &-&-&  MS & I                        & U N N N       &b \\
004.0$-$02.6 & PHR 1804$-$2645 &0.62&   24 $\times$ 45& Y &$\sim$90&A/90& CB &  G                        & N N N Y       &b \\
\hline                                                                                 
144.8$+$65.8 & BE UMa          &2.29& 180 & Y  &84 &A/85& LB &  -                        &  U U U U  &j,n\\
055.4$+$16.0 & A 46            &0.47& 53 $\times$ 70& Y  &80 &A/80& LB &  I                      & U U U U  &b,c,g\\
053.8$-$03.0 & A 63            &0.46& 42 $\times$ 290& Y  &88 &A/90& CB & J                        & U U U U  &g \\
136.3$+$05.5 & HFG 1           &0.58& 540& N  &28 &-&  Irr &I,L                      &  U U U U &f\\
253.5$+$10.7 & K 1-2           &0.68& 50 $\times$ 130 & N  &$\sim$45 &A/45&  LB & L,J                       & U U U U &b,i,m\\
283.9$+$09.7 & DS 1            &0.36& 290 $\times$ 335& N  &62&-& PB & I,L                      &  U U U U &b \\
329.0$+$01.9 & Sp 1            &2.91& 120 & N  &-&A/5& LB &  G,R               &  Y \textbf{Y} Y \textbf{Y} &c \\
349.3$-$01.1 & NGC 6337        &0.17& 60 $\times$100& N  &9&A/10& LB &  G,L,J,R                        &  Y \textbf{Y} Y \textbf{Y}    & l \\
005.1$-$08.9 & Hf 2-2          &0.40& 25 & N  &-&-&  MS & L,R                       &  U U U U &b,d \\
017.3$-$21.9 & A 65            &1.00& 220 $\times$ 240& N  &-&A/10& LB & J                      &  U U U U &e,h\\
009.6$+$10.5 & A 41            &0.23& 15 $\times$ 22& N  &66 & A/70 & CB &  R                        & U U U U &b,g \\
335.2$-$03.6 & HaTr 4          &1.74& 24 $\times$ 30 & N  &-&B/45 & PB &  -                        &  U U U U &c \\
215.6$+$03.6 & NGC 2346       &15.99& 58 $\times$ 172 & N  &50&A/50& CB & G,R                        &  Y Y \textbf{Y} \textbf{Y} & c \\
135.9$+$55.9 & SBS 1150$+$599A &0.16& 4 & N  &-&-& Irr & L?                       &  \textbf{Y} \textbf{Y} Y Y & k\\
341.6$+$13.7 & NGC 6026        &0.53& 60& N  &82&-& Irr & -                     &  U U U U &d \\
      \hline
   \end{tabular}
   \begin{flushleft}
      \emph{Morphologies:} CB: Canonical bipolar; LB: Likely bipolar; PB: Possible bipolar; MS: Multiple shells; Irr: Irregular \\
      \emph{Morphological Traits:} G: Gatley's rule satisfied; I: ISM interaction; L: Low-ionisation knots or filaments; J: Jets; R: Equatorial ring\\
      \emph{IRAC:} U: Unavailable; Y: Nebular detection; N: Nebular non-detection.\\ 
      \emph{References:} (a) Paper I, (b) this work, (c) Bond \& Livio (1990), (d) Schwarz, Corradi \& Melnick (1992), (e) Walsh \& Walton (1996),\\ (f) Heckathorn, Fesen \& Gull (1982), (g) Pollacco \& Bell (1997), (h) Hua, Dopita \& Martinis (1998), (i) Corradi et al. (1999), (j) Bond (2000),\\ (k) De Marco (2009), (l) Corradi et al. (2000), (m) Exter, Pollacco \& Bell (2003), (n) Liebert et al. (1995)  
   \end{flushleft}
\end{table*}

\subsection{Narrow-band optical imaging}

As part of the acquisition process during our Gemini South programs we took advantage of the exceptional imaging capabilities provided by GMOS (Hook et al. 2004). Table \ref{tab:gmos_log} records the filters and exposure times of images used in this work. The central CCD was read out with 2 $\times$ 2 binning to give a 5.5 $\times$ 2.5 arcmin field-of-view sampled at 0.145\arcsec/pixel. The central wavelength and FWHM of the filters are 499.0/4.5 nm (OIII), 514.0/8.8 nm (OIIIC), 656.0/7.2 nm (Ha), 662.0/7.1 nm (HaC) and 672.0/4.4 nm (SII). The Ha filter includes H$\alpha$ and both [NII] lines, while the HaC filter effectively acts as an [NII] $\lambda$6584 filter. 

\begin{table}
   \centering
   \caption{GMOS acquisition images used in this work.}
   \label{tab:gmos_log}
   \begin{tabular}{llrllr}
      \hline\hline
      Name & Filter & Exp. & Name & Filter & Exp. \\
      \hline
      Pe 1-9 & Ha & 180        &  K 6-34 & Ha & 240 \\
             & OIII   & 120    &         & OIII & 240 \\
             & OIIIC   & 120   &         & OIIIC & 120 \\
      Sab 41 & OIII & 240      &  PHR 1756$-$3342 & Ha & 300\\
             & Ha & 240        &                  & OIIIC & 120\\
             & HaC & 120       &  PHR 1757$-$2824 & Ha & 150\\
      M 3-16 & Ha & 120        &                  & HaC & 60\\
             & OIII & 120      &  PPA 1759$-$2834 & OIII & 300\\
      H 2-29 & Ha & 60         &                   & g$'$ & 90\\
             & SII & 60        &  BMP 1800$-$3408 & Ha & 300\\
             & OIII & 60       &                  & r$'$         & 90\\
      M 2-19 & OIII & 120      & PHR 1804$-$2913 & Ha &  60\\
             & Ha & 120        &                & HaC & 60\\
             & SII & 120       &                &     &   \\
      \hline
   \end{tabular}
\end{table}

\begin{figure}
   \begin{center}
      \includegraphics[scale=1.0]{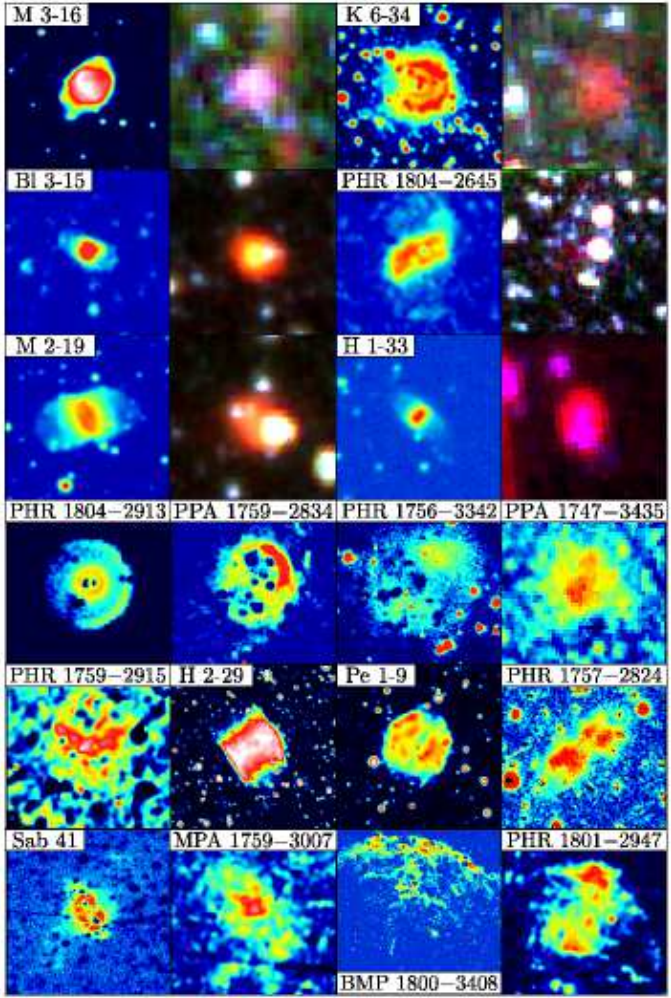}
   \end{center}
   \caption{A montage of mostly H$\alpha$+[NII] false-colour images and \emph{Spitzer} IRAC colour-composite images of the 18 nebulae selected from Paper I for morphological analysis. 
   }
   \label{fig:morphs}
\end{figure}

Figure \ref{fig:morphs} presents a montage of mostly H$\alpha$+[NII] images of the OGLE sample (Tab. \ref{tab:morph}). The main contribution comes from our Gemini programs, followed by the ESO NTT/EMMI programs 079.D-0764(B) and 67.D-0527(A) for H 1-33 and Bl 3-15, respectively (see Paper I), and the AAO/UKST SuperCOSMOS H$\alpha$ survey (SHS, Parker et al. 2005). We further add \emph{Spitzer} colour-composite images described in Sec. \ref{sec:mir}. Image dimensions are all 30 $\times$ 30 arcsec$^2$ except for PHR 1804$-$2913 (15 $\times$ 15 arcsec$^2$), PHR 1804$-$2645, MPA 1759$-$3007, BMP 1800$-$3408, PHR 1801$-$2947 and PHR 1759$-$2915 (1 $\times$ 1 arcmin$^2$), and the SHS quotient image of Sab 41 (4 $\times$ 4 arcmin$^2$). 
Throughout this work all figures are presented in the standard orientation unless otherwise indicated (North is up and East is to left) and each image is centred on the central star.
The SHS images are also complemented later by broad-band $B_J$ images from the SuperCOSMOS Sky Surveys (SSS, Hambly et al. 2001) in constructing colour-composite images from SHS H$\alpha$ (red), SHS Short-Red (green) and SSS $B_J$ (blue).

Our primary focus in this work is on the new OGLE sample rather than previously well-studied known sample. 
Section \ref{sec:morph} will present a selection of more detailed images of the OGLE sample than that provided in Fig. \ref{fig:morphs}. A reappraisal of available images for certain members of the known sample will also be made alongside the OGLE sample, however for those not covered we refer the reader to the final column of Tab. \ref{tab:morph}. 

\subsection{Gatley's rule and mid-infrared imaging}
\label{sec:mir}
As a complement to narrow-band optical imagery, near- and mid-infrared imagery can help reveal an intrinsic bipolar morphology regardless of the projected 2D morphology. This development started with Zuckerman \& Gatley (1988) and was later solidified by Kastner et al. (1994, 1996) who conducted narrow-band near-infrared (NIR) imaging surveys of PNe in the 2.122 $\mu$m H$_2$ emission line. They found a strong propensity for canonical bipolar PNe to display H$_2$ emission in their waist regions. Excluding canonical bipolar PNe, H$_2$ detections were found \emph{only} amongst torus or ring-like structures interpreted to be inclined equatorial waists of bipolar nebulae. This is known as \emph{Gatley's rule}, namely that the detection of H$_2$ emission is a unique marker of a canonical bipolar PN (Kastner et al. 1996). 

While we acknowledge that Gatley's rule has yet to be rigorously tested with large and unbiased samples, there are some consistent observational and theoretical indications that the rule holds and is suitable for use in identifying bipolar PNe. An independent study by Guerrero et al. (2000) found definite H$_2$ detections in 13/15 bipolar PNe. These encouraging results are supported theoretically by Aleman \& Gruenwald (2004) who modelled the physical conditions responsible for the presence of H$_2$ in PNe. They found higher H$_2$ mass fractions are associated with higher central star temperatures, which is consistent with the higher than average temperatures amongst bipolar PNe (Corradi \& Schwarz 1995), thereby providing a simple physical explanation of Gatley's rule.

Although we lack narrow-band near-infrared H$_2$ imaging for our sample, we can make use of the next best thing, observations made with the Infrared Array Camera (IRAC; Fazio et al. 2004) onboard the \emph{Spitzer Space Telescope}. The four IRAC bands are centred on 3.6 $\mu$m (band 1), 4.5 $\mu$m (band 2), 5.8 $\mu$m (band 3) and 8.0 $\mu$m (band 4). Band 4 in particular serves as a good tracer of H$_2$ emission containing two H$_2$ lines and an [Ar III] line (Hora et al. 2004). Bands 2 and 3 are also sensitive to H$_2$ emission but not as much as band 4. It follows that a strong band 4 detection and moderate to absent detections in the other bands satisfies Gatley's rule, i.e. the PN is highly likely to have an intrinisc bipolar morphology. 
Emission lines from atomic transitions and PAHs do occur within the broad IRAC bands however (e.g. Cohen et al. 2007), and while their application in this context is not free from ambiguity, it nonetheless serves as a useful tool in the absence of proper narrow-band H$_2$ imaging. Indeed, Kwok et al. (2008) found excellent agreement between a narrow-band H$_2$ image of NGC 6583 and the \emph{Spitzer} IRAC band 4 image while demonstrating the bipolar (double-coned) morphology of NGC 6583. 
Another dramatic case where H$_2$ emission unambiguously identifies bipolar morphology not necessarily evident in optical emission lines can be seen in the velocity integrated H$_2$ image of NGC 7027 (Cox et al. 2002). 

We have searched the \emph{Spitzer} science archive using the \textsc{leopard} tool for publicly available IRAC observations for our sample. Tab. \ref{tab:morph} records the status of each IRAC band as either being unavailable (U), a nebular detection (Y) or a non-detection (N). Almost all observations were conducted as part of the ongoing GLIMPSE surveys (Benjamin et al. 2003; Churchwell et al. 2009). Since GLIMPSE is confined mostly to $|b|\lesssim1\degr$, much of the known sample at higher latitudes have no available images. 
Images of SBS 1150$+$599A were taken serendipitously during observations of the blue compact dwarf galaxy SBS 1150$+$599 (Wu et al. 2006), whereas NGC 2346 was observed as part of program ID 68 (PI: Fazio). 
All colour-composite \emph{Spitzer} images presented in Fig. \ref{fig:morphs} and elsewhere in this paper were constructed using IRAC bands 4 (red), 3 (green) and 2 (blue). In these images a ruddish or red-orange hue, expected to coincide with the waist of a bipolar PN, satisifies Gatley's rule. 
Non-detections for the faintest MASH PNe could be explained by their very low surface brightness.

\section{Classification Scheme}
\label{sec:scheme}
\subsection{Geometric models}
\label{sec:models}

We have constructed two rudimentary geometric models for a range of inclination angles to compare qualitatively with the projected 2D morphologies of our sample. Columns seven and eight of Tab. \ref{tab:morph} encapsulate these comparisons in the rough choice of model and inclination, and the overall morphology, respectively. Depending on the inclination of the system and which geometric model is used, the overall morphology may be one of canonical bipolar (CB), likely bipolar (LB) or possible bipolar (PB). Nebulae showing multiple-shell (MS) or irregular (Irr) morphologies are classified separately without the models. The morphological traits listed in column eight of Tab. \ref{tab:morph} show whether Gatley's rule is satisfied (G, see Sec. \ref{sec:mir}), ISM interaction is likely (I), or an equatorial ring is present (R). Low-ionisation structures (L) and jets (J) are defined in Sec. \ref{sec:lis} and discussed further in Sec. \ref{sec:originlis}.

\begin{figure*}
   \begin{center}
      \large{Model A}\\
\includegraphics[scale=0.35]{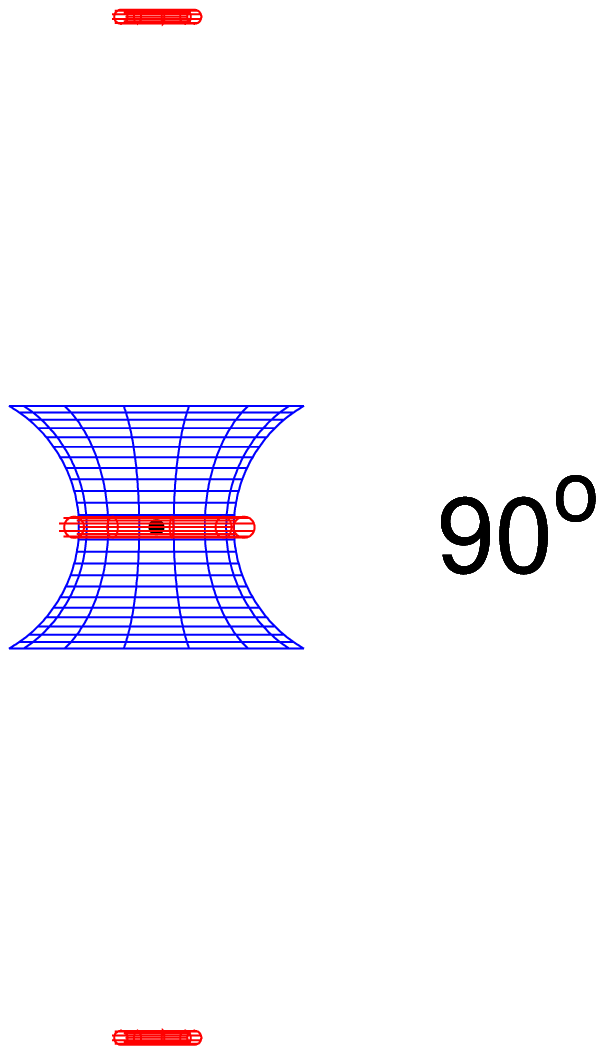}
\hspace{-0.50cm}
\includegraphics[scale=0.35]{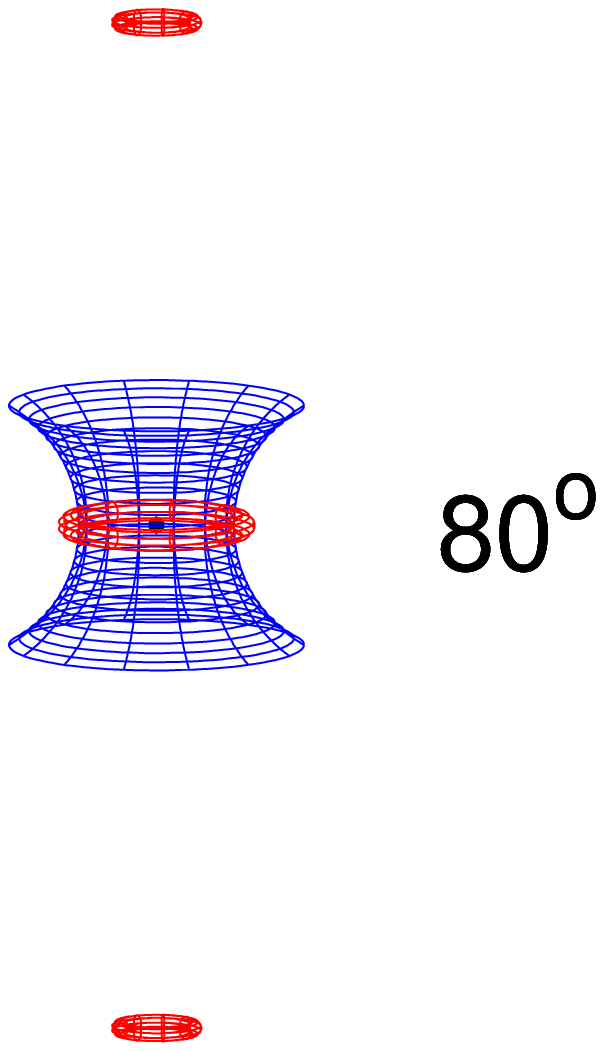}
\hspace{-0.50cm}
\includegraphics[scale=0.35]{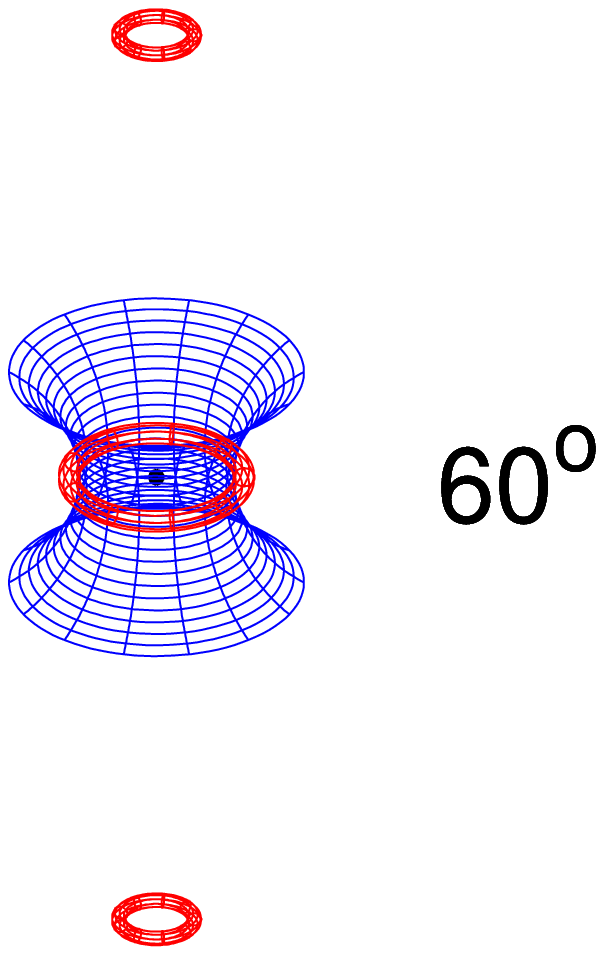}
\hspace{-0.50cm}
\includegraphics[scale=0.35]{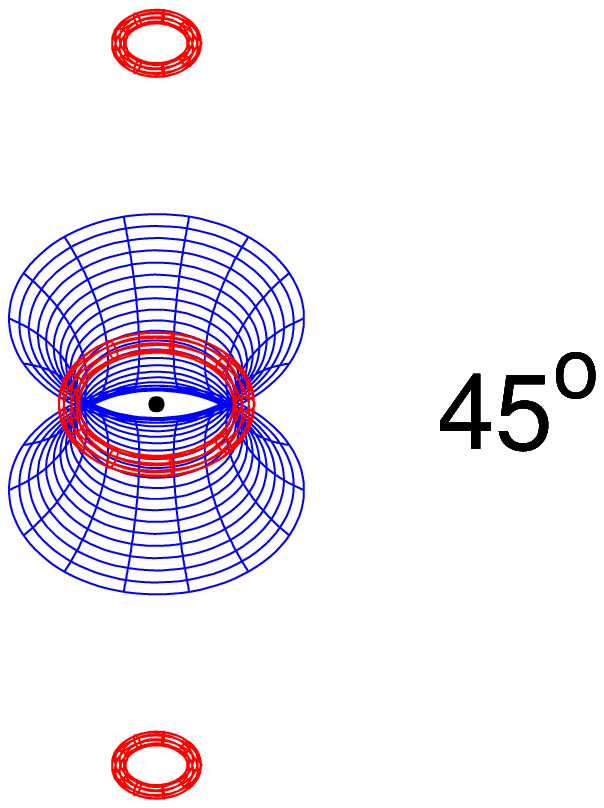}\\
\vspace{-1cm}
\includegraphics[scale=0.35]{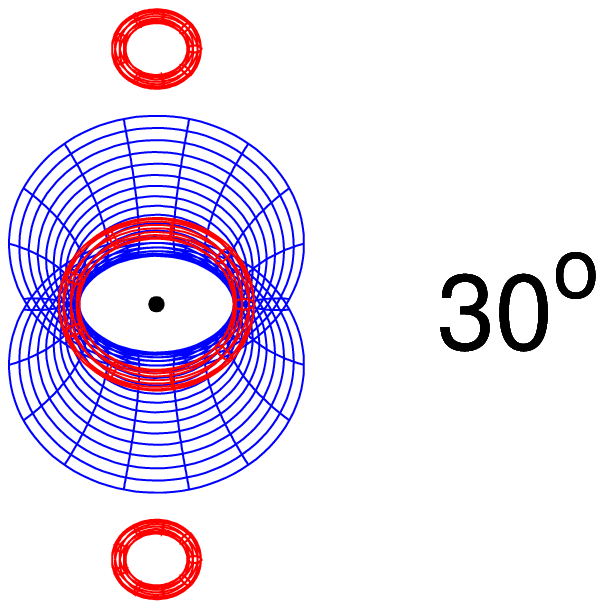}
\hspace{-0.50cm}
\includegraphics[scale=0.35]{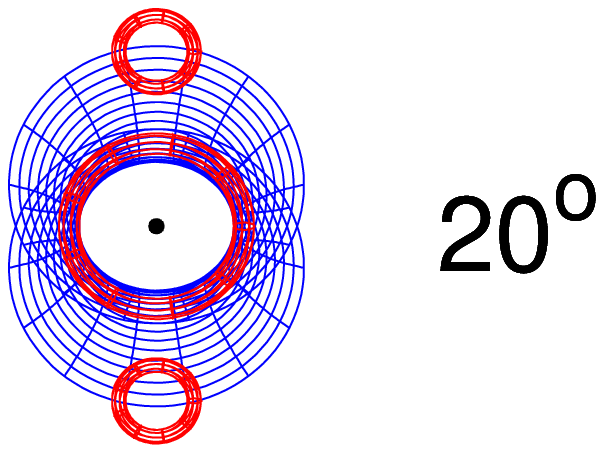}
\hspace{-0.50cm}
\includegraphics[scale=0.35]{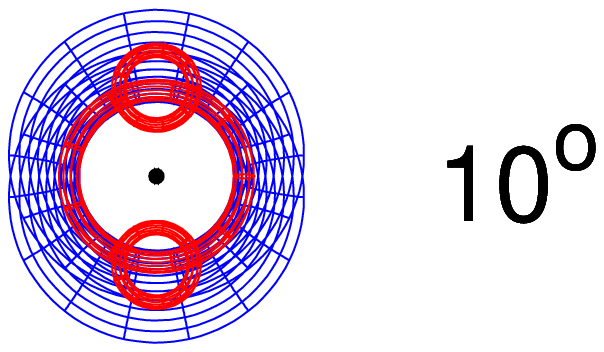}
\hspace{-0.50cm}
\includegraphics[scale=0.35]{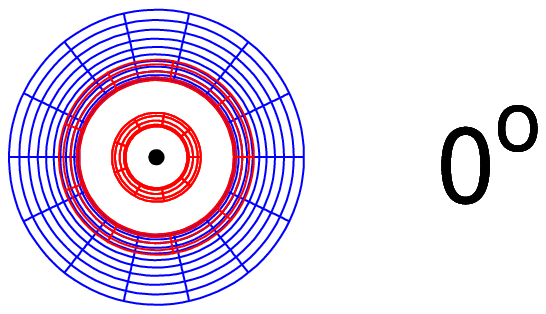}\\
\vspace{-1.0cm}
     \large{Model B}\\
\vspace{-0.85cm}
\includegraphics[scale=0.35]{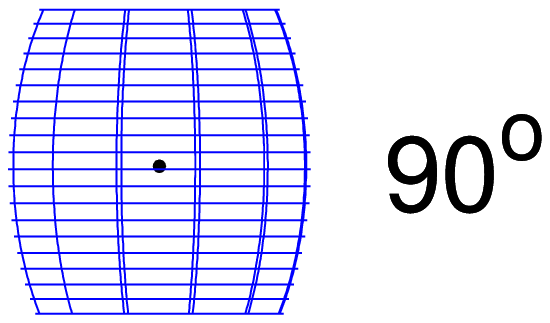}
\hspace{-0.50cm}
\includegraphics[scale=0.35]{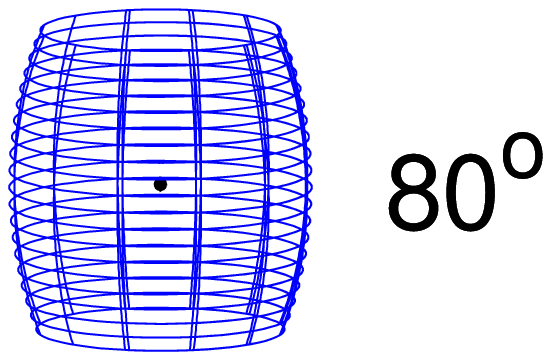}
\hspace{-0.50cm}
\includegraphics[scale=0.35]{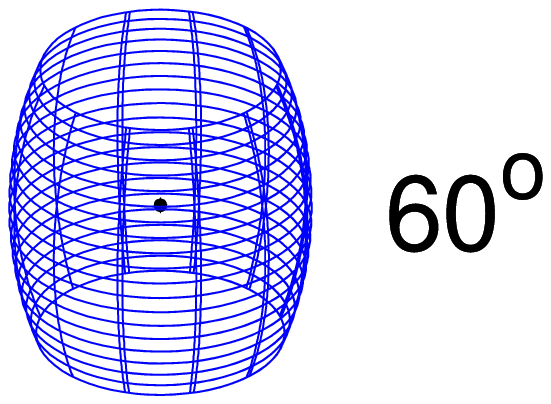}
\hspace{-0.50cm}
\includegraphics[scale=0.35]{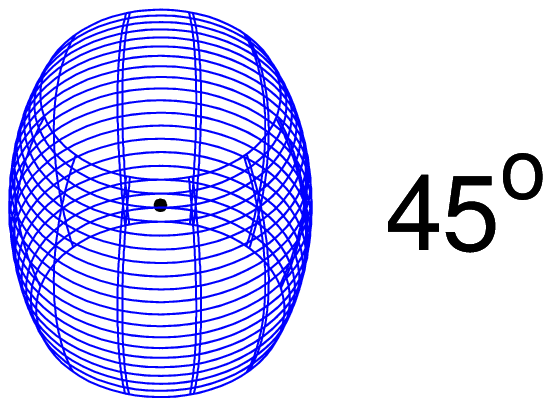}\\
\vspace{-2cm}
\includegraphics[scale=0.35]{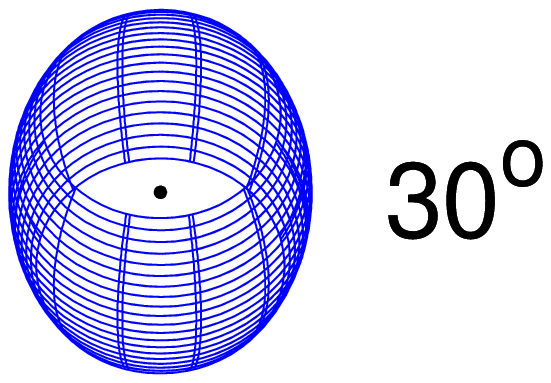}
\hspace{-0.50cm}
\includegraphics[scale=0.35]{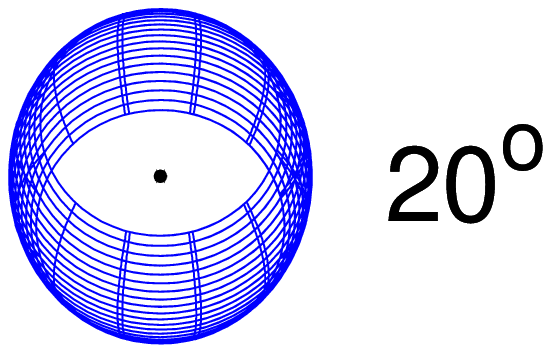}
\hspace{-0.50cm}
\includegraphics[scale=0.35]{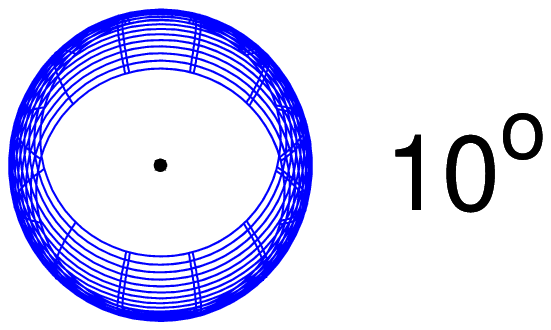}
\hspace{-0.50cm}
\includegraphics[scale=0.35]{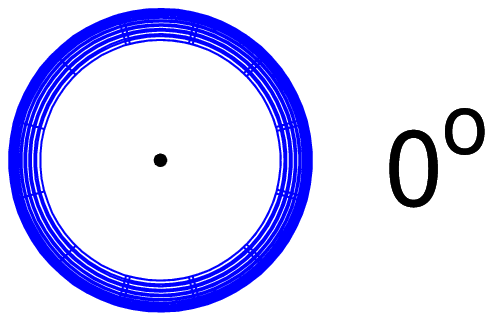}
\vspace{-1.5cm}
   \end{center}
   \caption{Model A is a simple bipolar adapated from Pollacco \& Bell (1997). Model B is a rudimentary adaptation of the Frank et al. (1993) model. The inclination angle of the nebulae range from 90$\degr$ (edge-on) to 0$\degr$ (pole-on).} 
   \label{fig:model}
\end{figure*}

Figure \ref{fig:model} presents our slight extension of the Pollacco \& Bell (1997) model (hereafter, Model A) which has been used successfully to reveal the intrinsic bipolar morphology of A 63 (Mitchell et al. 2007). The components are a main outflow (blue hyperboloid), an equatorial ring of low-ionisation material (large red torus), polar ejecta or jets (small red torii) and a central star (black dot). Note at low inclination angles the projected radial distance of the jets depends heavily on their height above the equatorial plane. We do not demand a precise match to Model A since the aspect ratios or even overall shape of the main outflow are considered to be quite flexible. Other shapes may be incorporated including a `peanut' or double cone (Kwok et al. 2008) structure, but these are less common amongst the observed morphologies. Matches to Model A are typically given the likely bipolar (LB) classification or alternatively canonical bipolar (CB) if they meet the strict criteria outlined in Sec. \ref{sec:canonical}.

Also included in Fig. \ref{fig:model} is a more rudimentary version of the Frank et al. (1993) model (hereafter, `Model B'). Model B consists of a main outflow (blue modified prolate spheroid) and a central star (black dot), but it could conceivably have the same low-ionisation ring or jets as seen in Model A. 
The formation of Model B involves a much lower density contrast between the equatorial and polar directions than Model A.
It would therefore be considered more elliptical than bipolar, especially since it lacks the conventional `pinched' or concave waist. This convention however does not account for evolutionary and inclination effects befitting bipolar nebulae (Icke, Preston \& Balick 1989). Model B would therefore also describe `early' bipolar PNe (e.g. NGC 6886, Balick \& Frank 2002; see also Hajian et al. 2007) which have a barrel-shaped or convex waist against the established convention. In such cases the polar outflows are less developed and are yet to create the effect of a pinched waist as seen in more evolved nebulae (e.g. NGC 7026, Balick \& Frank 2002). While these outflows could simply be added to Fig. \ref{fig:model}, we have omitted them to emphasise that even some more evolved nebulae will have lost their polar outflows to leave a `bipolar core'. In most cases Model B nebulae are classified possible bipolar (PB) since an elliptical model can reproduce the same morphology. If polar outflows are also present then a likely bipolar (LB) classification is assigned.

Multiple-shell (MS) nebulae are also identified in our sample (e.g. Chu, Jacoby \& Arendt 1987). A separate model is not required for these nebulae which are discussed later in terms of features drawn from both models (see Sec. \ref{sec:ms}). 
Less certain cases may be classified as irregular (Irr) typically when there are insufficient morphological details available to facilitate a clear classification. 

Even though we lack detailed kinematic information for each nebula, there is much to be gained from qualitative analysis of high-quality images within the framework of these models. Apart from the identification of unambiguous morphologies and an overall picture of the sample, the models can be applied to similar nebulae to unlock a common intrinsic morphology difficult to infer otherwise. This approach is also highly complementary to kinematic modelling that sometimes require similar interpretation to work around limitations inherent in the method (e.g. low-inclination and low-velocity situations). Before conducting our qualitative study we first describe our nomenclature regarding low-ionisation structures belonging to Model A.

\subsection{Low-ionisation structures}
\label{sec:lis}

At least 17\% of PNe exhibit intricate low-ionisation structures (LIS) that are often considerably smaller than the overall nebula diameter (Corradi et al. 1996). Narrow-band [NII] or [SII] images are useful in revealing LIS, but they are best isolated by constructing an H$\alpha$+[NII]/[OIII] image that removes density variations to highlight just changes in ionisation.
Gon{\c c}alves et al. (2001) reviewed PNe exhibiting LIS, their kinematics and their purported formation mechanisms (see also Gon{\c c}alves 2004). 

Studies of LIS have not been helped by a largely inconsistent approach to their nomenclature in the literature. Perhaps the most ambiguous case being the highly collimated `jet-like' pairs often quoted as `jets' (e.g. Exter et al. 2003) even though their velocities not substantially different from their nebulae (Gon{\c c}alves 2004). We describe \emph{knots} as unresolved or small LIS with an aspect ratio of unity, and \emph{filaments} as \emph{knots} with an aspect ratio larger than unity (Gon{\c c}alves et al. 2001). 
We adopt a stricter definition of \emph{jets} to be opposing polar outflows located well outside the main nebula and with typically lower surface brightness than equatorial filaments (e.g. Mitchell et al. 2007; small red torii in Model A). Jets may appear to be located within the radius of the main nebula because of projection effects, with their radial position dependent on the inclination of the nebula and the height of the jets from the equatorial plane. We avoid the `jet-like' nomenclature often applied to highly collimated equatorial filaments that have no relation to our definition of jets.

\section{Nebular Morphologies} 
\label{sec:morph}
\subsection{Canonical bipolars}
\label{sec:canonical}

The \emph{canonical bipolars} are described by Corradi \& Schwarz (1995) to have a well-defined equatorial waist or ring-like structure from which two fainter, extended lobes diverge in the polar direction. As a group the majority of them are evolved and so are best represented by Model A or some slight variation of Model A at intermediate--high inclination. At inclinations lower than $i\sim45\degr$ kinematic modelling is required to confirm an intrinsic bipolar morphology. When the kinematics exist for low-inclination nebulae confirming bipolarity we retain a likely bipolar morphology (LB) such that our canonical bipolar (CB) classification is compatible with the literature. 

Amongst the known sample only NGC 2346 (Bond \& Livio 1990) and A 63 (Mitchell et al. 2007) have been identified as canonical bipolar nebulae with inclinations greater than $i\sim45\degr$. Such a small fraction is puzzling considering the expectation of a high density contrast between the equatorial and polar directions for post-CE nebulae (Sandquist et al. 1998). 
This landscape has now changed dramatically with the more representative OGLE sample containing at least four canonical bipolars, all of which have significantly shorter periods than NGC 2346. These are described below in addition to A 41 which upon reappraisal exhibits a canonical bipolar morphology. 
Further bipolars may be found in Sec. \ref{sec:likely} and Sec. \ref{sec:possible}, but many of these require full kinematic modelling.

\subsubsection{M 2-19, H 2-29 and A 41}

\begin{figure}
   \begin{center}
      \includegraphics[scale=1.00]{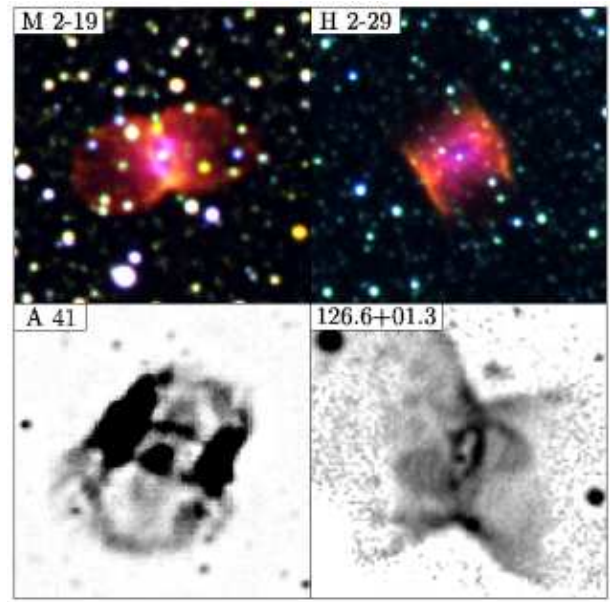}
   \end{center}
   \caption{Gemini GMOS colour-composite images of M 2-19 and H 2-29 made from H$\alpha$+[NII] (red), [SII] (green) and [OIII] (blue) reveal equatorial rings. Rings are also seen in reproduced H$\alpha$+[NII] images of A 41 (Pollacco \& Bell 1997) and PN G126.6$+$01.3 (Mampaso et al. 2006). Image dimensions are 30 $\times$ 30 arcsec$^2$ and some unsharp masking has been applied to all except H 2-29.
   }
   \label{fig:rings}
\end{figure}

Figure \ref{fig:rings} depicts Gemini GMOS colour-composite images of M 2-19 and H 2-29 alongside reproduced H$\alpha$+[NII] images of A 41 (Pollacco \& Bell 1997) and PN G126.6$+$01.3 (Mampaso et al. 2006). Apart from NGC 2346, the most canonical bipolar morphology of all post-CE nebulae is demonstrated by M 2-19. It best matches Model A at $i\sim45\degr$ and as expected satisfies Gatley's rule (Fig. \ref{fig:morphs}). 
A prominent equatorial ring is seen amongst all nebulae in Fig. \ref{fig:rings} and this is complemented by a separate inner nebula tilted to the major axis in all except H 2-29. 
Both A 41 and M 2-19 are the first confirmed post-CE nebulae to show this quadrupolar trait consistent with precession associated with a binary nucleus (Manchado et al. 1996; Mampaso et al. 2006). A binary nucleus appears to be present in PN G126.6$+$01.3, but the definitive proof of an orbital period has not yet been realised. In M 2-19 the feature is only visible in the unsharp masked [OIII] image as included in the colour-composite. Inspection of Model A reveals an inclination effect to be likely responsible for the less-developed lobes of the eclipsing H 2-29 compared to M 2-19 and A 41. 
Even more evolved than M 2-19 and A 41 is PN G126.6$+$01.3 which exhibits jets extending two arcminutes either side of the central star (Mampaso et al. 2006).
H 2-29 also shows some tentative evidence for a polar outflow in the SW direction (Fig. \ref{fig:morphs}) as well as LIS particularly along its edges.

\subsubsection{H 1-33, PHR 1757$-$2824, PHR 1759$-$2915 and PHR 1804$-$2645}

The remaining canonical bipolars are both highly collimated (H 1-33 and PHR 1757$-$2824) and less collimated (PHR 1759$-$2915 and PHR 1804$-$2645) nebulae (Fig. \ref{fig:morphs}). Gatley's rule applies as expected for H 1-33 and PHR 1804$-$2645 with band 4 detections in the waist region of both nebulae. All objects exhibit eclipses except for H 1-33 whose inclination may be similar to M 2-19 at $i\sim45\degr$. 
A bipolar waist is apparent in both PHR 1759$-$2915 and PHR 1804$-$2645 along PA$\sim90\degr$ from which very faint lobes diverge north and south. This is less evident in PHR 1759$-$2915 which seems to be a fainter version of PHR 1709$-$3931 (Parker et al. 2006). 

\subsection{Likely bipolars}
\label{sec:likely}
\subsubsection{Sab 41}
\label{sec:sab41}
\begin{figure}
   \begin{center}
      \includegraphics[scale=1.000]{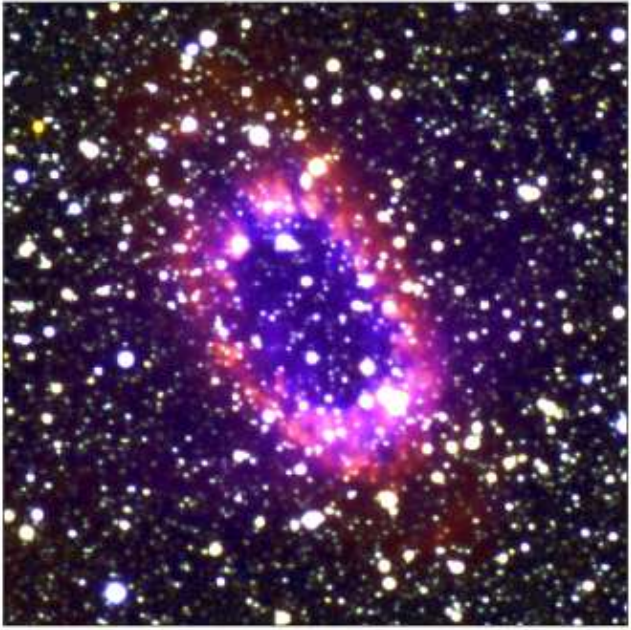}
   \end{center}
   \caption{
   Gemini GMOS colour-composite 1.7 $\times$ 1.7 arcmin$^2$ image of Sab 41. Colours are H$\alpha$ and [NII] (red), [NII] (green) and [OIII] (blue).
   }
   \label{fig:sab41_ring}
\end{figure}

\begin{figure}
   \begin{center}
   \includegraphics[angle=270,scale=1.0]{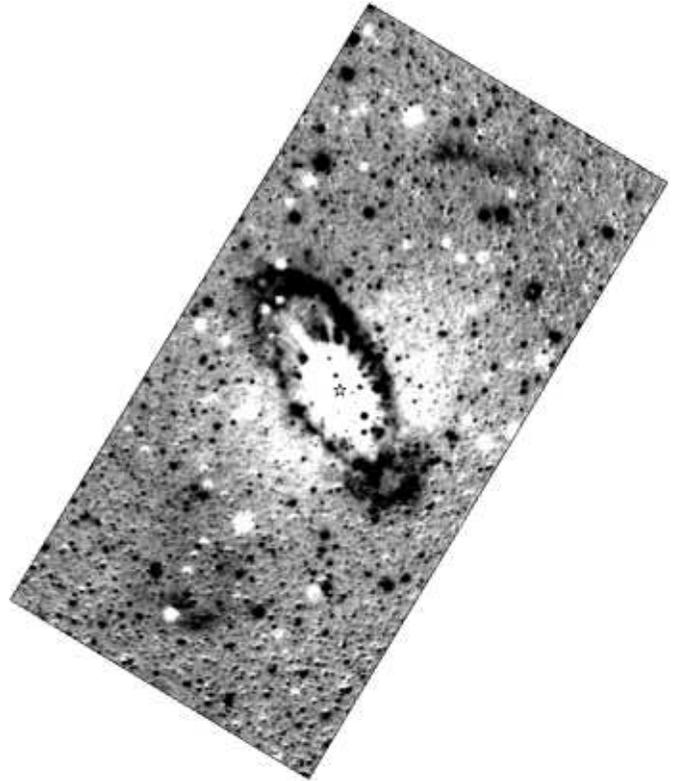}
   \end{center}
   \caption{Gemini GMOS image of Sab 41 constructed by dividing [OIII] by H$\alpha$+[NII].
   Note the opposing faint [NII]-rich end-caps 1.55 arcmin from the central star marked in this 4 $\times$ 2 arcmin$^2$ image.
   }
   \label{fig:sab41_large}
\end{figure}

Discovered by Zanin et al. (1997), Sab 41 was first imaged by Cappellaro et al. (2001) to reveal an elliptical nebula measuring 42\arcsec\ $\times$ 77\arcsec. It was rediscovered independently by Parker et al. (2006) as PHR 1748$-$3538 who noticed faint extensions beyond the main nebula in the considerably deeper SHS image (Fig. \ref{fig:morphs}). Figures \ref{fig:sab41_ring} and \ref{fig:sab41_large} present our GMOS images which provide a much more complete view of the nebula and its extensions. The main elliptical ring measures 30\arcsec\ $\times$ 50\arcsec in [OIII] and slightly larger in [NII] at 30\arcsec\ $\times$90\arcsec. The [NII] ring and in particular its terminating SW end is seen amongst bipolar nebulae like Lo 18 (Schwarz et al. 1992), while the [OIII] emission traces the main nebula body. Inside the elliptical ring are several radially distributed low-ionisation filaments with extended tails pointing away from the central star. 
Also visible are polar jets that end in shocked [NII] emission 1.55 arcmin from the central star just like those seen in A 63 (Mitchell et al. 2007). In the plane of the sky these are bent 30$\degr$ away from the major axis defined by the [OIII] outflow. All these features make Sab 41 an exemplary fit to Model A at $i\sim40\degr$.

\subsubsection{NGC 6337 and Sp 1}

\begin{figure}
   \begin{center}
      \includegraphics[scale=1.00]{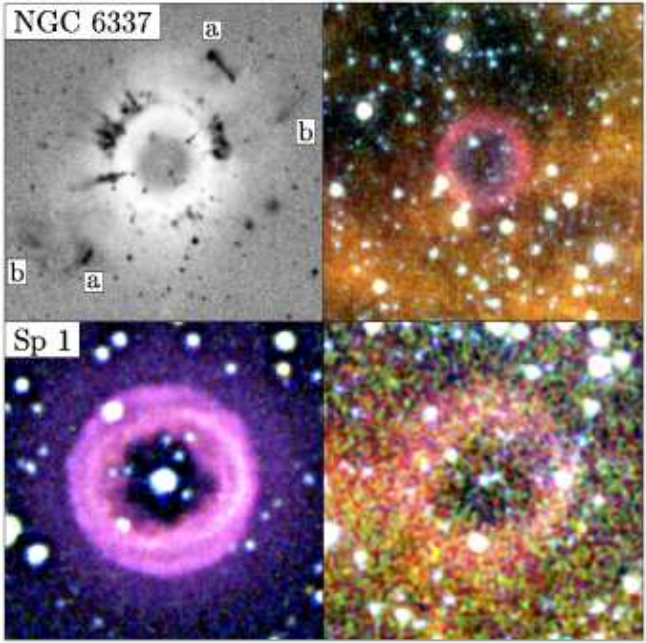}
   \end{center}
   \caption{The low inclination bipolars NGC 6337 and Sp 1. The first column shows the [OIII]/[NII] image of NGC 6337 (Corradi et al. 2000) and an unsharp masked SHS/SSS colour-composite image of Sp 1. These are accompanied by their \emph{Spitzer} IRAC colour-composite counterparts in the second column. Image dimensions are 2.5 (NGC 6337) and 2.0 (Sp 1) arcmin a side.
   }
   \label{fig:largepoleon}
\end{figure}

Although NGC 6337 is a relatively new discovery amongst the known sample (Hillwig 2004; Hillwig, Bond \& Af{\v s}ar 2006), previous interest in its low-ionisation structures gave rise to the first kinematic study of the nebula (Corradi et al. 1996, 2000). Figure \ref{fig:largepoleon} reveals these low-ionisation structures in an [OIII]/[NII] image reconstructed from archival data (Corradi et al. 2000). 
Corradi et al. (2000) found only a small difference in the measured radial velocity either side of the main ring and interpreted this as evidence for a pole-on bipolar nebula with low-ionisation filaments radially distributed around a projected equatorial density enhancement. A low-inclination was vindicated by lightcurve analysis which gave a best-fit inclination of $i=9\degr$ (Hillwig et al. 2006), while a bipolar nebula is supported by Gatley's rule being satisfied by a ring of \emph{Spitzer} IRAC band 4 emission (Fig. \ref{fig:largepoleon}). 

Having established NGC 6337 as a pole-on bipolar befitting Model A at $i=10\degr$, explaining the pairs of low-ionisation features outside the main nebula body is now straightforward. Kinematic studies tell us the motion of pair `a' is slow and pair `b' is fast (Corradi et al. 2000; Garc\'ia-D\'iaz et al. 2009). Both studies support a polar outflow interpretation for pair `b' given its high velocity, however pair `a' is more difficult to accurately model given its combination of low-inclination and low-velocity. Garc\'ia-D\'iaz et al. (2009) recognised the inability of their model to constrain the distance of pair `a' from the equatorial torus, so it is still plausible that pair `a' represents a slow-moving polar outflow. 
The modelling effort is further complicated by a high likelihood that both outflows are bent, not only to each other, but also to the polar axis. Nevertheless, the identification of essentially identical polar outflows in A 63 and Sab 41 (small red torii built into Model A) lead us to strongly suggest pair `a' is indeed a similar polar outflow. The strong resemblance between Sab 41 (Fig. \ref{fig:sab41_large}) and NGC 6337 is discussed in Sec. \ref{sec:originlis}. 

Bond \& Livio (1990) similarly suspected Sp 1 (Fig. \ref{fig:largepoleon}) to be a pole-on bipolar because of its ridged appearance. A kinematic study has apparently confirmed this classification (Mitchell, priv. communication to Zijlstra 2007) and this is supported again by Gatley's rule being satisfied by a ring of \emph{Spitzer} IRAC band 4 emission as in NGC 6337 (Fig. \ref{fig:largepoleon}).

\subsubsection{A 65 and PPA 1759$-$2834}
\label{sec:a65}
\begin{figure}
   \begin{center}
      \includegraphics[scale=0.65]{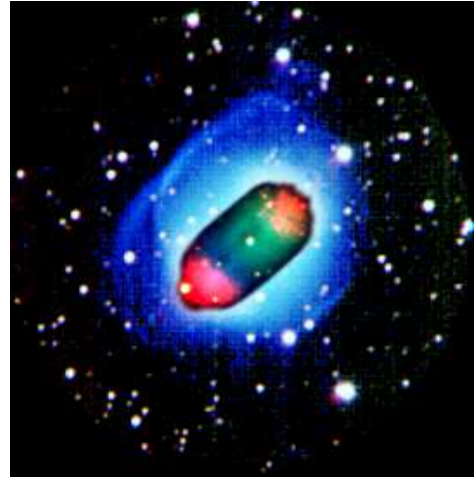}
   \end{center}
   \caption{A colour-composite image of A 65 constructed using deep [NII] (red), H$\alpha$ (green) and [OIII] (blue) images taken from Hua et al. (1998). 
   }
   \label{fig:a65}
\end{figure}

Abell 65 has traditionally been classified as an elliptical PN (Bond \& Livio 1990; Pollacco \& Bell 1997), however this classification did not take into account deeper [OIII] images that reveal considerably more outer structure (Walsh \& Walton 1996; Hua, Dopita \& Martinis 1998). Figure \ref{fig:a65} reproduces the Hua et al. (1998) images as a colour-composite image with the inner region shown at lower contrast overlaid on the full nebula at higher contrast. 
Walsh \& Walton (1996) suggested the dark lane seen along the major axis of the bright elliptical component was actually a gap similar to what we see in Model A for $i\sim45\degr$ aligned along PA$\sim45\degr$, leading to A 65 being classified as a bipolar waist if only the shallow images are examined (e.g. De Marco 2009). However, adopting this model does not fit the observed outer lobes of A 65, nor does it explain the large [NII] regions at either end of the main body. Furthermore, the central star is not placed in the centre of the gap (which itself is rather irregular), whereas one may expect from the model a more centred position and a sharper, more defined gap. Goldman et al. (2004) discuss similar dark patches in NGC 1360 and attribute their origin to either lower gas densities or higher extinction. We reject the latter possibility based on unpublished \emph{Spitzer} IRAC photometry conducted under program ID 40020 (PI: Fazio) which shows the dark patches of NGC 1360 are still present in the mid infra-red. This must also be the case for A 65 since Walsh \& Walton (1996) found no change in H$\alpha$/H$\beta$ across its dark lane.

Taking a fresh approach a good fit to the outer lobes and lemon shaped cross-section seems to be Model A at $i\sim10\degr$ and PA$\sim135\degr$. This interpretation accommodates the [NII] regions as polar outflows which may otherwise be difficult to explain. At lower contrast the [NII] regions seem fragmented and this, combined with their conical shape, appears consistent with simulations of a ballistic (polar) outflow (e.g. Fig. 6 of Raga et al. 2007). 
Higher resolution images and kinematics of the [NII] regions are needed to confirm this hypothesis.

Another case in point for deep, multiple emission line imaging is PPA 1759$-$2834. In Paper I the low-resolution SHS image shows a round nebula, however the GMOS [OIII] image reveals a dramatically different picture (Fig. \ref{fig:morphs}). A round inner nebula is resolved to have internal structure suggestive of a Model B fit at $i\sim45\degr$, while two faint extensions are seen in the polar direction which are absent in the H$\alpha$+[NII] image (just as for A 65). These extensions lead us to classify the nebula as likely bipolar despite the strong Model B resemblance. A brightened NE rim is probably caused by ISM interaction, and while this could be an alternative explanation for the extensions (e.g. HFG 1, Sec. \ref{sec:hfg1}), this is unlikely since their position is not coincident with the brightened NE rim.

\subsubsection{K 1-2}
\begin{figure}
   \begin{center}
      \includegraphics[scale=1.0]{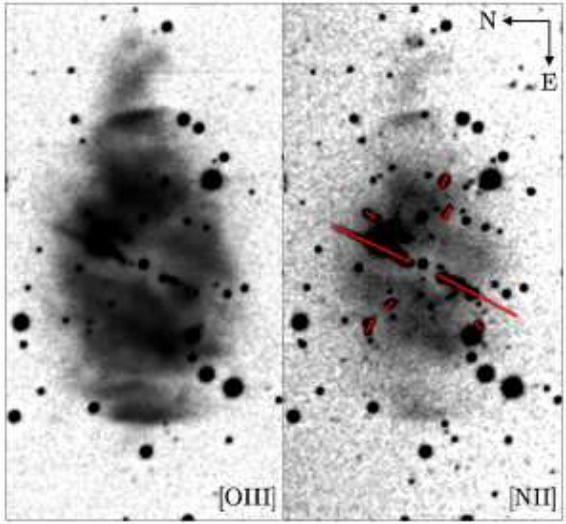}
   \end{center}
   \caption{Enhanced [OIII] and [NII] images of K 1-2 measuring 90 $\times$ 170 arcsec$^2$. Low-ionisation filaments are traced by red lines.}
   \label{fig:k1-2}
\end{figure}

All previous studies of K 1-2 have focused solely on the multiple opposing pairs of low-ionisation filaments (Corradi et al. 1999; Exter et al. 2003), rather than the overall nebula structure. Figure \ref{fig:k1-2} reproduces the ESO NTT/EMMI [OIII] and [NII] images of K 1-2 from ESO program ID 55.D-0550 (Exter et al. 2003) which have been enhanced to reveal the outer structures. The orientation of the images was chosen for best comparison with Model A which is a very good match at the inclination of $i$$=$40--50$\degr$ derived from analysis of the nucleus (De Marco et al. 2008). The splayed nature of the multiple opposing filament pairs (marked in Fig. \ref{fig:k1-2}) seem to trace out the equatorial ring as in Model A. Low filament velocities common to both K 1-2 and NGC 6337 (Corradi et al. 1999, 2000; Gon{\c c}alves 2004) further support this interpretation. In the central region an apparent inner gap and elongation in the N-S direction closely mirrors the elliptical shape and cross-section of the gap in Model A. We interpret the ionisation structure of the `ears' or `multiple-shells' referred to by previous authors as shocked outflows similar to IC 4634 and M 3-16 (Raga et al. 2008; Sec. \ref{sec:m3-16}).

\subsubsection{A 46} 
\begin{figure}
   \begin{center}
    \includegraphics[scale=0.475]{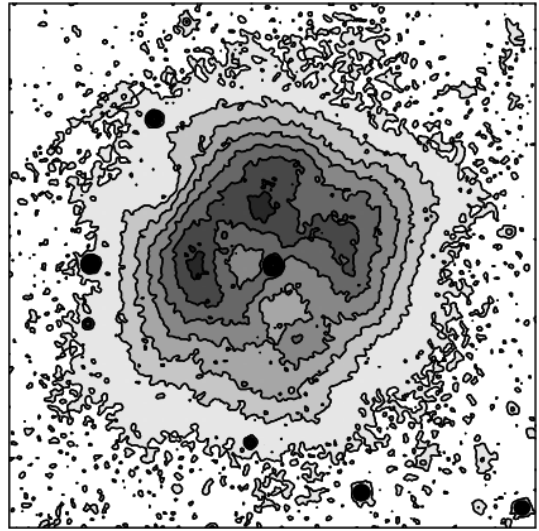}
   \end{center}
   \caption{Contoured WHT Taurus-2 [OIII] 2 $\times$ 2 arcmin$^2$ image of A 46.}
   \label{fig:a46}
\end{figure}

With its eclipsing nucleus, comparison of A 46 with Model A should be similar to A 63. Figure \ref{fig:a46} presents a previously unpublished WHT Taurus-2 30 minute [OIII] exposure of A 46 taken on 18 May 1995. Although of similar depth to the Pollacco \& Bell (1997) image, the WHT image enables clarification of some features identified by them. The inner contours best fit Model A if we align the polar axis with PA$\sim$55$\degr$, however it is the outer contours that are of most interest. Pollacco \& Bell (1997) claimed a `bow-shock' feature towards the NE, however we only find an indentation. We attribute this feature to probable ISM interaction where the concave entry (NE) and convex exit (SW) points mark the path of the ISM flowing through the nebula (e.g. Fig. 1 of Dgani \& Soker 1998). This is consistent with the lower surface brightness SW lobe (Bond \& Livio 1990) and probably explains the bridge-like feature remarked by Pollacco \& Bell (1997) at PA$\sim$135 as just a contrast difference.

\subsubsection{BE UMa, Bl 3-15 and MPA 1759$-$3007}
BE UMa appears to be clearly bipolar with an equatorial ring and polar outflows (Bond 2000). However, if we adopt Model A at the inclination of the system derived from the eclipsing nucleus (84$\degr$), then the apparent inclination of the equatorial ring seems too low. This may be explained by its more evolved and therefore distended nature and we await a more detailed kinematic study of the nebula. We suspect Bl 3-15 to be a bipolar PN complete with probable jets as in A 63 that befit Model A. An unresolved nebular core does not allow us to narrow down an inclination and while one could argue for an elliptical morphology, the satisfaction of Gatley's rule (Fig. \ref{fig:morphs}) supports our bipolar classification. 
Similarly, for MPA 1759$-$3007 a central overdensity and probable bipolar lobes are present, suggesting Model A is a plausible fit for an inclination angle of $i\sim55\degr$, but higher quality images are required before we can classify the nebula as a canonical bipolar.

\subsection{Multiple shells}
\label{sec:ms}

In a dramatic departure from Bond \& Livio (1990), we have identified four PNe with multiple-shell morphologies (MSPNe), namely PHR 1804$-$2913, Pe 1-9, K 6-34 and Hf 2-2. MSPNe exhibit considerable morphological (Chu, Jacoby \& Arendt 1987) and kinematic (Guerrero, Villaver \& Manchado 1998) diversity which naturally gave rise to multiple proposed formation models (Chu et al. 1987; Chu 1989). Bond \& Livio (1990) favoured multiple mass ejections on the tip of the AGB to explain the absence of MSPNe in their sample because the CE phase would interrupt such a scenario. We now know that the simplest double-shell PN requires only a combination of photo-ionisation and wind interaction (Sch\"onberner et al. 1997; Corradi et al. 2003) and this favours the existence of at least some MSPNe with binary nuclei.

\subsubsection{PHR 1804$-$2913}
PHR 1804$-$2913 is unique amongst our sample with a round double-shell morphology (Fig. \ref{fig:morphs}). Measuring 8\arcsec\ across the outer shell is more than twice the 3.5\arcsec\ diameter of the inner shell and the western side appears to be brightened by interaction with the ISM. The eastern edge of the outer shell has an intensity 30--45\%  that of the inner shell, far too bright to be a true halo which would be considerably fainter (Corradi et al. 2003). We measured a similar intensity ratio for the outer shell of M 2-2 from the H$\alpha$ IAC catalogue image (Manchado et al. 1996a). Unlike M 2-2, there is no evidence for LIS in the outer shell of PHR 1804$-$2913. 

\subsubsection{Pe 1-9, K 6-34 and Hf 2-2}
\begin{figure}
   \begin{center}
      \includegraphics[scale=1.00]{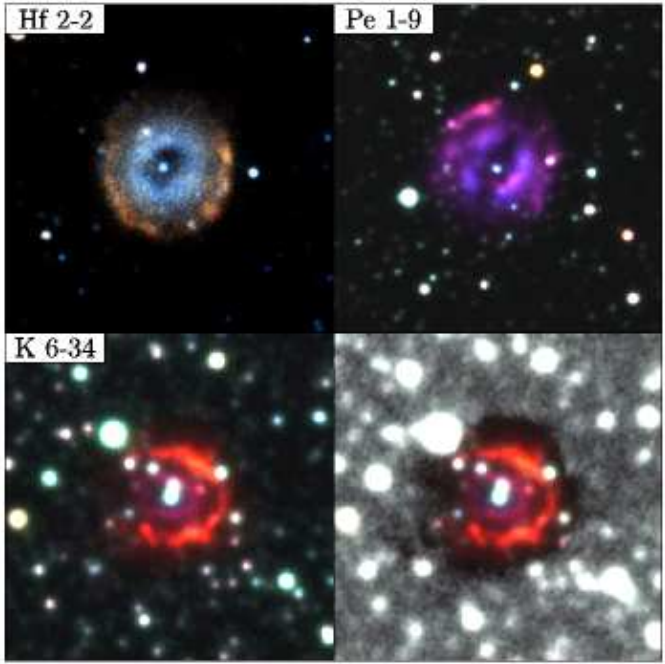}
   \end{center}
   \caption{Multiple-shell nebulae Hf 2-2, Pe 1-9 and K 6-34. 
   The PNIC colour-composite image of Hf 2-2 (Schwarz et al. 1992) is constructed from H$\alpha$ (red), [OIII] (blue) and their sum (green). Gemini GMOS colour-composite images of Pe 1-9 and K 6-34 were made using Ha (red), OIIIC (green) and OIII (blue) exposures. The second K 6-34 panel features an Ha$-$OIIIC image that reveals faint [NII] emission features outside the main nebula (overlaid).
   Image dimensions are 45 $\times$ 45 arcsec$^2$ (Hf 2-2) and 30 $\times$ 30 arcsec$^2$ (Pe 1-9 and K 6-34).
  }
   \label{fig:poleon}
\end{figure}

The remaining MSPNe Pe 1-9, K 6-34 and Hf 2-2 are depicted in Fig. \ref{fig:poleon}. 
Overall this group is quite homogeneous with an inner shell dominated by [OIII] emission surrounded by an outer partial shell of low-ionisation knots and small filaments. Unlike Hf 2-2 and Pe 1-9, the [OIII] emission in K 6-34 does not extend to the outer shell. This suggests a different morphology for K 6-34, namely a likely bipolar given the satisfaction of Gatley's rule and a probable jet revealed by [NII] emission surrounding the nebula similar to NGC 6337. 

The O'Dell, Weiner \& Chu (1990) model for NGC 2392 seems most suitable for the sample. 
This model is equivalent to adding the low-ionisation components of Model A to Model B. 
Hf 2-2 and K 6-34 are probably viewed close to pole-on, while reconciling the edge-on Pe 1-9 is a little more difficult since the main nebular body is split into two halves. It may be that Pe 1-9 is just the inner shell of the O'Dell et al. (1990) model viewed edge-on without the equatorial LIS. In that case the LIS seen in the rim of Pe 1-9 have just migrated from the inner shell, except perhaps for one bent filament immediately SW of the CSPN. 
Although their prolate ellipsoidal cores do not make them bipolar, the MSPNe appear to at least be axisymmetric and they do exhibit LIS consistent with the other post-CE nebulae. 
Many other PNe show the same ionisation structure including for example NGC 7662 (Balick 1987, 1998), M 2-2 (Manchado et al. 1996a) and NGC 2392 (Balick 1987). Whether LIS in this configuration are a reliable indicator of CE evolution is discussed in Sec. \ref{sec:enuclei}.

\subsection{Possible bipolars}
\label{sec:possible}
\subsubsection{DS 1}
\begin{figure}
   \begin{center}
      \includegraphics[scale=1.00]{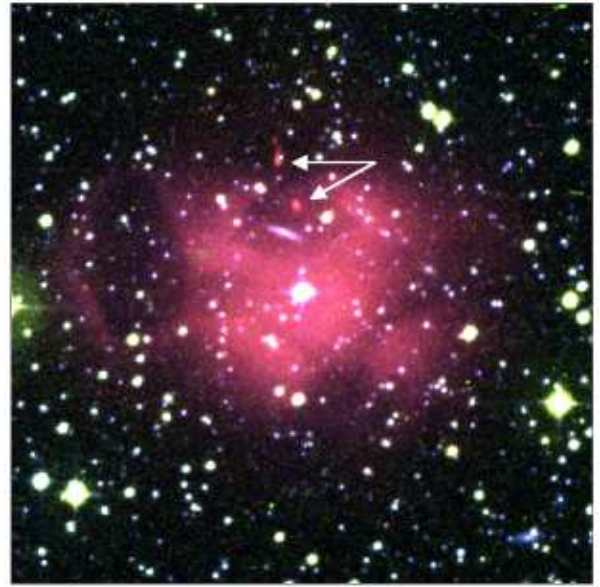}
   \end{center}
   \caption{An SHS/SSS colour-composite 7 $\times$ 7 arcmin$^2$ image of DS 1 showing two newly identified LIS north of the central star (arrowed). 
   }
   \label{fig:ds1}
\end{figure}

In Fig. \ref{fig:ds1} we identify two northern LIS in DS 1 for the first time. The outer most appears more collimated with a trail of smaller components while the other is more like a knot. The knot was missed by Bond \& Livio (1990) due to their lower quality image where it appears stellar, while the northern more collimated LIS is located outside their image. These LIS are real because they appear in both the H$\alpha$ and B$_J$ images. We speculate that the eastern loop of DS 1 represents a polar outflow, placing the major axis of the nebula at PA $=$ 90$\degr$ and the LIS are therefore consistent with being in the equatorial plane. The overall morphology however remains elusive, possibly bipolar, and is probably affected by density effects (e.g A 65). 

\subsubsection{HaTr 4, PHR 1756$-$3342 and PHR 1801$-$2947}
 
HaTr 4 (Bond \& Livio 1990) appears to be best described by Model B at $i\sim45\degr$. The low density contrast required to generate Model B requires us to classify HaTr 4 as only a possible bipolar. Alternatively, HaTr 4 may resemble H 2-29 but the close match with Model B seems more convincing and the original undersampled lightcurve means that it may or may not be eclipsing (De Marco et al. 2008).

Both PHR 1756$-$3342 and PHR 1801$-$2947 exhibit rather elliptical nebulae but on closer inspection both have two opposing overdensities (Fig. \ref{fig:morphs}). They may be represented by Model B at $i\sim30\degr$ and we classify them as possible bipolars for now.

\subsection{Irregulars}
\subsubsection{M 3-16}
\label{sec:m3-16}

\begin{figure}
   \begin{center}
      \includegraphics[scale=1.0]{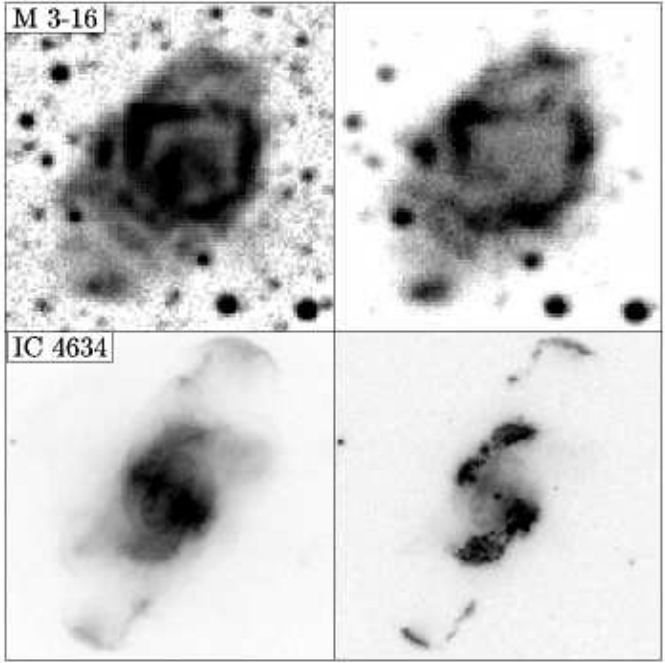}
   \end{center}
   \caption{Gemini GMOS images of M 3-16 and archival \emph{HST} images of IC 4634 show many similarities in both [OIII] (first column) and [NII] (second column). The [NII] image of M 3-16 was created in the same way as Fig. \ref{fig:sab41_large}. Image dimensions are 15 $\times$ 15 arcsec$^2$ (M 3-16) and 20 $\times$ 20 arcsec$^2$ (IC 4634).
   }
   \label{fig:m3-16}
\end{figure}

The intrinsic morphology of M 3-16 is difficult to discern but there are a number of similarities that can be drawn with its closest analogue, the point-symmetric IC 4634 (Guerrero et al. 2008). Figure \ref{fig:m3-16} displays GMOS images of M 3-16 alongside archival \emph{HST} images of IC 4634. The [OIII] emission traces the overall nebular morphology and especially the inner most ellipsoidal nebula aligned with the major axis at PA$\sim$135$\degr$. Note the inner nebula largely disappears in [NII] and if a classification were to be made elliptical might be the most suitable for both nebulae. This is supported by an unusual violet \emph{Spitzer} false-colour (Fig. \ref{fig:morphs}) that mitigates against a bipolar classification using Gatley's rule since a number of different scenarios can produce this false-colour (Cohen et al. 2007). 

Large outflows at the ends of the major axis are just visible in the [OIII] image of each nebula, but they become sharper in [NII]. These features are well-reproduced by models as fast-moving photo-ionised neutral clumps (Raga et al. 2008). The [NII] image of M 3-16 also reveals low-ionisation filaments and some knots which do not display the same high degree of point symmetry as in IC 4634. The eclipsing nucleus of M 3-16 tells us the outflows are indeed in the polar direction, which is probably also the case for IC 4634. The inclination of IC 4634 appears to be low and so in a similar case to NGC 6337 the position of the jets is not well constrained by the kinematics and they could well be polar outflows (Guerrero et al. 2008). If the inclination of IC 4634 were increased to match M 3-16 we might suspect its LIS to `fold in' on themselves to appear similar to those in M 3-16. 

\subsubsection{HFG 1}
\label{sec:hfg1}
 
In the discovery images (Heckathorn, Fesen \& Gull 1982) then nebula of HFG 1 apparently resembles the multiple-shell PNe IC 1295 (Manchado et al. 1996a) and Wray 17-1 (Corradi et al. 1999). A more favourable explanation of the brightened SE rim is provided by an ISM interaction caused by a high proper motion. This was first suggested by Xilouris et al. (1996) and was largely confirmed by Boumis et al. (2009) who elucidated a 20\arcmin\ long tail of emission consistent with the CSPN proper motion via deep imaging. The overall morphology is then indeterminate having been substantially modified by ISM interaction. A low-ionisation filament was first identified and studied in the discovery paper, although no kinematic information was given.

\subsubsection{SBS 1150$+$599A}
De Marco (2009) reproduced an \emph{HST} F658N image of SBS 1150$+$599A that shows a round nebula with two collimated filaments. It is peculiar that spectroscopic studies of the nebula (Tovmassian et al. 2001; Jacoby et al. 2002) have not recorded any low-ionisation species that could be attributed to these filaments, especially since the F658N filter is centred on [NII] $\lambda$6584 with only $\sim$3\% transmission at H$\alpha$. Attributing the filaments to H$\alpha$ emission alone is therefore unlikely and we suspect further IFU or high spatial resolution longslit observations may recover [NII] emission. The filaments therefore would not be representative of the intrinsic morphology which is round or unresolved.
It is interesting to note the \emph{Spitzer} IRAC colour-composite image has the same violet hue as M 3-16 so we cannot classify the nebula as bipolar using Gatley's rule.

\subsubsection{BMP 1800$-$3408, PPA 1747$-$3435 and NGC 6026}
 
Our deep GMOS image of BMP 1800$-$3408 reveals many filaments arranged in an asymmetric shell quite reminiscent of a supernova remnant (Fig. \ref{fig:morphs}). 
Yet the hot star confirmed by our GMOS spectrum has the same radial velocity as the nebula and so is certainly the ionising source of the nebula. Further study is required to establish whether this is a genuine evolved PN.
The nebulae of PPA 1747$-$3435 and NGC 6026 (Schwarz et al. 1992) are too diffuse to confidently identify any features for morphological classification. High quality GMOS images of PPA 1747$-$3435 were taken but did not improve over the SHS images given the diffuse nature of the nebula. 

\section{Discussion}
\subsection{A penchant for bipolarity}
\label{sec:penchant}
 
We are now able to assess the fraction of bipolars amongst our post-CE sample (Tab. \ref{tab:morph}). 
Removal of the as yet unconfirmed PHR 1759$-$2915, MPA 1759$-$3007 and PHR 1801$-$2947 leaves us with 30 nebulae, of which there are eight canonical bipolars (27\%). Accounting for inclination effects and incorporating Gatley's rule brings an additional 9--10 likely bipolars that contribute to a plausible bipolar fraction of $\sim$60\%. This bipolar fraction now includes the two essentially confirmed bipolars NGC 6337 and Sp 1 that we left as likely bipolar for consistency with the literature.
It may even be as high as 65--70\% if some of the possible bipolars are included.
Around 10\% of the sample are MSPNe, some of which may turn out to be likely bipolars viewed pole-on (e.g. K 6-34 which satisfies Gatley's rule). Detailed studies of the irregular nebulae that make up the remaining 20\% of the sample are needed to resolve their status.

Our qualitative analysis shows for the first time a penchant for bipolarity exists amongst the post-CE nebulae. 
This is a major step forward from the paradigm where the unusual NGC 2346 was considered to be the only bipolar (Soker 1997, 1998). It seems highly likely that CE evolution produces bipolar nebulae more often than single progenitors, but future morphological studies of nebulae (in particular those around single CSPN) are required to build a control sample for a definitive answer. The first point of call for such a study would be those nebulae with extensive time-series photometry that rule out a close binary. This would include the pulsators and non-detections in Ciardullo \& Bond (1996), as well as the non close-binary CSPN with `true' identifications from sufficiently sampled fields in Paper I. Even if high-quality images were available for the latter sample, there may be many members distorting the results which have either very late-type main sequence or planet-sized companions below the detection limit of current surveys or which may have undergone a previous CE interaction (Sec. \ref{sec:enuclei}). Another option would be to refer to large morphological studies of Galactic PNe as a reference point, but since these do not adequately treat inclination effects any comparison here would be premature. 

The 30\% of our sample with canonical bipolar morphologies have considerably shorter orbital periods than NGC 2346. Soker (1997, 1998) proposed that NGC 2346 avoided the CE phase for as long as possible during its evolution, but clearly the new canonical bipolars are free of this restriction. Indeed, the more numerous and representative OGLE sample fulfils the greater parameter space, whereas NGC 2346 occupies only a narrow evolutionary space (Soker 2002). Our results are therefore consistent with recent predictions, but they still invalidate the first critical observations of earlier works (Soker 1997, 1998). 

Another application of the canonical bipolars in the OGLE sample concerns the sample selection for Tables 3 and 4 of Soker (1997). In general, the former contains bipolars with pronounced lobes and waists (e.g. M 2-19), while the latter contains more `boxy' nebulae reminiscent of bipolar waists but without pronounced lobes (e.g. H 2-29). We've highlighted M 2-19 and H 2-29 to demonstrate again the influence of inclination on what may otherwise be construed \emph{solely} as evolutionary effects. Consider Model A (Fig. \ref{fig:model}), those nebulae viewed close to edge-on (H 2-29) will often appear more `boxy' or compact than those viewed at intermediate inclinations (M 2-19) despite them both having essentially the same intrinsic morphology.

\subsection{A binary formation scenario for low-ionisation structures} 
\label{sec:originlis}

The origin of LIS is uncertain given the considerable diversity in their observed properties (Gon{\c c}alves et al. 2001), not to mention the sometimes confusing nomenclature (Sec. \ref{sec:lis}). An accretion disk established via binary interaction has been proposed to power jets (Soker \& Livio 1994; Nordhaus \& Blackman 2006), however binary formation scenarios are yet to be explored for other LIS. In what follows we propose a binary scenario in general terms after discussing the occurence of LIS in our morphological study. 

Table \ref{tab:morph} summarises the LIS identified during the morphological study according to Sec. \ref{sec:lis}, namely whether filaments and knots are present in the equatorial plane, or if low-surface brightness jets were ejected in the polar direction. Our approach to nomenclature was motivated by the strong resemblance between NGC 6337 and Sab 41, which together bridge the divide between the jets of A 63 (Pollacco \& Bell 1997; Mitchell et al. 2007) and the prominent filaments of some other post-CE nebulae (e.g. K 1-2). At least 10 nebulae have equatorial knots or filaments (33\%), around 20\% have jets, while both jets and equatorial LIS are present in only 10\% of cases. 
Combining the two types gives at least 40\% of post-CE which have some kind of LIS, more than double the estimated rate of occurrence amongst larger PNe samples (Corradi et al. 1996). Together with the emerging trends in the sample, e.g. NGC 6337 and Sab 41, the relatively high occurence tends to suggest a binary origin for LIS.

We propose that LIS are a natural consequence of a photo-ionising wind interacting with clumps of neutral material during the PN phase which were \emph{already} deposited during the CE interaction. Gon{\c c}alves et al. (2009) require LIS to be mostly a mix of dust and H$_2$ in order to resolve a large discrepancy in their observed densities with model expectations. This material is highly concentrated towards the orbital plane with only a small amount towards the polar direction (Sandquist et al. 1998). Add a photo-ionising wind to this mix and we have the appropriate conditions to reproduce the slow-moving equatorial filaments and knots in post-CE nebulae (e.g. Raga, Steffen \& Gonz\'alez 2005). Polar outflows are better described by fast-moving neutral material interacting again with a photo-ionising wind (Raga et al. 2007, 2008). Observationally K 1-2 and M 3-16 fit the model with the signature `reverse bow-shock' ionisation structure in their polar outflows (e.g. IC 4634, Raga et al. 2008). The counterparts in e.g. A 63 and Sab 41 seem more evolved, which may explain the lack of detected [OIII] emission in their outflows. 
Overall this picture is in good agreement with the low-ionisation filaments in the Helix whose H$_2$ cores seem to pre-date the PN phase (Matsuura et al. 2009). Much more work is required to develop the scenario further, e.g. incorporating the role of spiral shocks in the spatial distribution of LIS (Sandquist et al. 1998; Edgar et al. 2008). 

\subsection{Can low-ionisation structures be used to identify a previous common-envelope interaction?}
\label{sec:enuclei}

The sample of post-CE nebulae is still too small to deduce any robust trends in nebular morphology. All were discovered via arduous photometric monitoring campaigns which find only 10-20\% as close binaries (Paper I). Potentially the elevated occurence of LIS amongst post-CE nebulae may serve as a useful tool in identifying other post-CE nebulae. Pre-selection for LIS could prioritise objects for photometric monitoring to quickly increase the sample of known binary CSPN. In the event where no close binary is found, then the nebula may still have been produced by CE evolution if we assume the secondary was destroyed. This is particularly applicable in the case of jets since they may be formed either via a stellar dynamo or via destruction of the secondary (Nordhaus \& Blackman 2006). 

We are planning photometric surveys of more biased samples dominated by LIS, but in the meantime we have surveyed the literature as part of a preliminary investigation. 
Since the Gon{\c c}alves et al. (2001) review, the number of PNe with LIS has grown considerably thanks to the PNIC that provides multiple emission line archival \emph{HST} and ground based images (e.g. Schwarz et al. 1992; Manchado et al. 1996). Particularly useful are the \emph{HST} F658 [NII] images used to measure nebular expansion rates with LIS (e.g. Hajian et al. 2007). We have identified $\sim$70 PNe mostly from the PNIC for our investigation, of which about half have CSPN classifications available. 

\begin{figure}
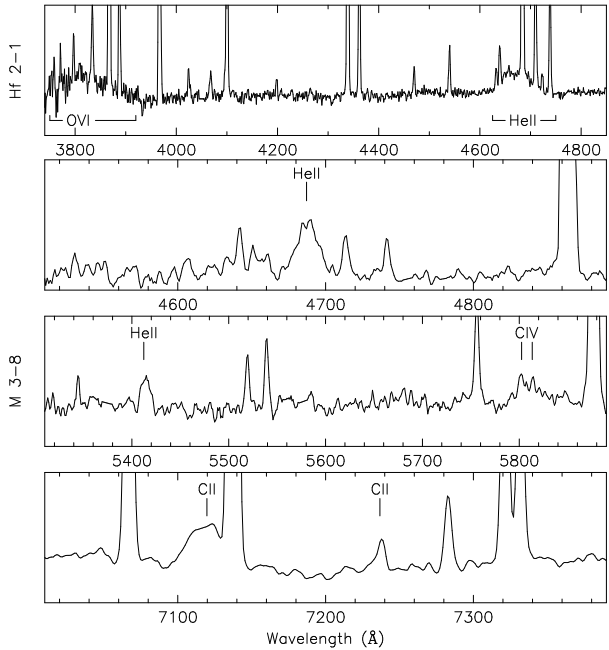

   \begin{center}
      \includegraphics[scale=0.35,angle=270]{hf2-1.ps}\\
      \includegraphics[scale=0.35,angle=270]{m3-8.ps}
   \end{center}
   \caption{Wolf-Rayet features in the AAOmega spectra of Hf 2-1 (top panel) and M 3-8 (bottom three panels).}
   \label{fig:wr}
\end{figure}

Table \ref{tab:cspnid} contains a summary of some of the better studied objects in the sample. The third column describes the LIS as jets (J), filaments (F) or both (JF), and filaments were chosen over jets for some objects where low velocities are known (Gon{\c c}alves 2004). For images we refer the reader to the PNIC for most nebulae and Corradi et al. (1999) for Wray 17-1. The fourth column shows the central star classification (M\'endez 1991; Werner, Heber \& Hunger 1991; Kaler, Stanghelli \& Shaw 1993; Acker \& Neiner 2003), while the last column indicates whether photometric variability has been reported or ruled out (Bond \& Grauer 1987; Bond \& Livio 1990; Ciardullo \& Bond 1996; Handler 2003). The variability described here is \emph{not from a close companion}, but rather reflects pulsations or variable winds. Ciardullo et al. (1999) found no wide companion in NGC 6826, NGC 6543, NGC 2371 and NGC 5882, and only a possible association in NGC 2392. Hf 2-1 is a new discovery made from our deep AAOmega (Sharp et al. 2006) observations described in Paper I (Fig. \ref{fig:wr}). 

\begin{table}
   \centering
   \caption{A select sample of PNe identified to have LIS.}
   \begin{tabular}{lllll}
      \hline\hline
      PN G & Name & LIS & CSPN & Variable\\
      \hline
      002.2$-$09.4 & Cn 1-5 & F & [WO4pec] & -\\
      003.1$+$02.9 & Hb 4 & JF & [WO3]     & -\\
      025.3$+$40.8 & IC 4593 & F & Of & Y \\
      029.2$-$05.9 & NGC 6751 & F & [WO4] & N\\
      083.5$+$12.7 & NGC 6826 & F & Of & Y\\
      096.4$+$29.2 & NGC 6543 & JF & Of-WR & Y \\
      189.1$+$19.8 & NGC 2371 & F & [WO] & Y \\
      197.8$+$17.3 & NGC 2392 & JF & Of & Y \\
      243.3$-$01.0 & NGC 2452 & F & [WO1] & N \\
      258.0$-$15.7 & Wray 17-1 & F & PG1159 & N\\
      307.2$-$03.4 & NGC 5189 & F & [WO1] & Y \\
      327.8$+$10.0 & NGC 5882 & JF& Of & -\\
      355.4$-$04.0 & Hf 2-1 & F & [WO] & N \\
      \hline
   \end{tabular}
   \label{tab:cspnid}
\end{table}

\begin{figure}
   \begin{center}
      \includegraphics[scale=1.00,angle=90]{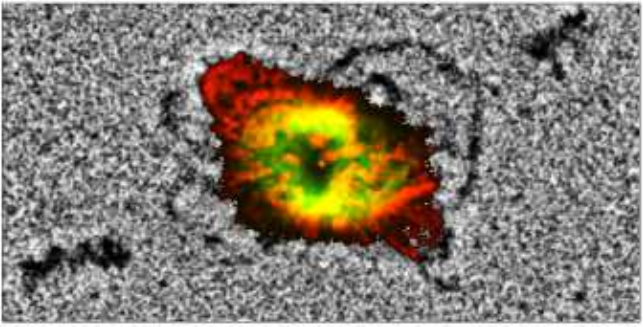}
   \end{center}
   \caption{\emph{HST} images of Hb 4 reveal jets and outer loops (enhanced H$\alpha$ background) in addition to many low-ionisation filaments in the central nebula overlaid as a colour-composite image of [NII] (red) and H$\alpha$ (green) emission. North is to the left and East is down in this 30 $\times$ 15 arcsec$^2$ image.}
   \label{fig:hb4}
\end{figure}

Our preliminary study shows the majority of the sample ($\sim$66\%) to have emission line nuclei of either Of ($\sim$33\%), [WR] (27\%) or PG1159 (6\%) type. Around 15\% are known or suspected binaries, while the remaining $\sim$18\% are non-emission line nuclei (some of which may turn out to be binary). Selecting for LIS therefore will give a mix of mostly binary and emission line nuclei which will require further observations to separate. Almost all the [WR] CSPN in the sample belong to the hot [WO] type that have more extreme and chaotic LIS covering their entire nebulae, presumably due to turbulence from the strong [WR] winds disrupting pre-existing LIS (Acker et al. 2002). The LIS in post-CE nebulae are much milder in comparison making the LIS in [WO] nebulae one of the few (if any) discriminating features between nebulae surrounding [WR] and non-[WR] PNe. Representative members of this group include NGC 6751, NGC 5189, NGC 2452 and Hb 4 (Fig. \ref{fig:hb4}). Our discovery of the [WO] nucleus of Hf 2-1 suggests HaTr 11 and Pe 1-17 may also turn out to be members of this class. 

If we assume LIS are indeed only produced by CE evolution, then the results of our preliminary study seem to be consistent with two speculative binary formation scenarios for Hydrogen-deficient central stars (De Marco \& Soker 2002; De Marco 2008). In short, these scenarios invoke either the destruction of a low-mass secondary or a final thermal pulse within the CE phase (causing a second CE phase) to enhance mass-loss to the point where the central star probably becomes Hydrogen-deficient. Such evolutionary paths are therefore highly likely to produce similar morphological traits amongst PNe surrounding single emission line nuclei. 

Unfortunately no direct evidence for a close binary with a [WR] component exists, leaving us dependent on only \emph{indirect} observational evidence. The scenarios provide a lever in that they make predictions about the dual-dust chemistry phenomenon (Cohen et al. 1999, 2002). The existence of dual-dust chemistry amongst only H-deficient CSPN is a critical outcome of the scenarios. Perea-Calder\'on et al. (2009) puts this link in doubt by their apparent discovery of some H-rich central stars with dual-dust chemistry. This finding however must be treated with caution since at least one object amongst their H-rich sample, M 3-8 (PN G358.2$+$04.2), is in fact H-deficient (Fig. \ref{fig:wr}). Deeper spectroscopy of the remaining H-rich dual-dust sample is required to settle their status and to further test the binary scenarios. It is also worth keeping in mind that the binary scenarios are still speculative and there may be another explanation for dual-dust chemistry that does not involve binary interactions (Perea-Calder\'on et al. 2009). Furthermore, Perea-Calder\'on et al. (2009) claim the binary fraction of 10--20\% (Paper I) is too low to support De Marco \& Soker (2002), however this is disputable given that in these scenarios the system would be undetectable because the secondary is likely to be destroyed. 

Assuming that the dual-dust predictions are correct, the strongest indirect evidence for the scenarios would be the simultaneous presence of strong post-CE morphological traits (jets and LIS) around a [WR] central star with dual-dust chemisty. We have identified Hb 4 as such an object with both prominent jets and low-ionisation structures (L\'opez, Steffen \& Meaburn 1997; Harrington \& Borkowski 2000) visible in archival \emph{HST} images (Fig. \ref{fig:hb4}). This is complemented by a [WO3] CSPN (Acker \& Neiner 2003) that was recently found to have dual-dust chemistry (Perea-Calder\'on et al. 2009). The low-ionisation filaments are consistent with the [WO] nucleus and the bent jets are quite strong evidence some binary interaction has occured (Soker \& Livio 1994). Independent time-series photometry of Hb 4 is required to definitively rule out a current binary as the OGLE-III data are most likely affected by nebular contamination (see Paper I and in particular Fig. 2 of Paper I). 

\section{Conclusion}
\label{sec:conclusion}
 
A core sample of 30 post-CE PNe was collated from the literature (15 PNe) and Paper I (15 PNe) to conduct a qualitative morphological study. High quality narrow-band optical imaging was sourced mainly from our GMOS programs on Gemini South, while some use was made of \emph{Spitzer} IRAC images to reveal H$_2$ emission in nebulae which is a strong indication of bipolar morphology (Gatley's rule). Morphological classifications were made with the assistance of geometric models adapated from Pollacco \& Bell (1997) and Frank et al. (1993) which assisted in accounting for the influence of inclination effects. 
Reappraisals of a selection of previously studied nebulae around binary CSPN yielded some new morphological details and helped frame the discoveries from Paper I. Spectroscopy of central stars in the OGLE sample will be presented in forthcoming papers in this series.

Our main conclusions are as follows:
\begin{itemize}
   \item Nearly 30\% of the sample are canonical bipolar nebulae, breaking the previous paradigm where
      NGC 2346 was the only post-CE canonical bipolar. This more representative sub-sample has much shorter periods than NGC 2346 demonstrating that it is indeed possible to produce a bipolar nebula without avoiding the CE phase until the final evolutionary stages (Soker 1997,1998). 
   \item A very plausible bipolar fraction of at least 60\% (possibly as high as 65-70\%) is reached once inclination effects and Gatley's rule are considered. This is the strongest indication yet that the morphologies of post-CE nebulae largely satisfy theoretical expectations of a high density contrast established during the CE phase. The link would be strengthened if reliable morphological studies of nebulae around non close-binary central stars were available. Such an ambitious study requires deep multiple emission line images, extensive kinematic modelling and consideration of inclination effects.
   \item No round, single-shell nebulae are found in the sample. Four multiple-shell nebulae come close, with the roundest being PHR 1804$-$2913, but at least one member of this sample may be a pole-on bipolar (K 6-34). The existence of these rounder nebulae may support the idea that the CE interaction can produce spherical PNe, but this can only be settled by dedicated monitoring surveys of round PNe. 
   \item Low-ionisation structures (LIS) appear to be quite common amongst post-CE nebulae (40\% occurence c.f. $\sim$17\% in general PN population). The emerging trends of nebulae with very similar morphological traits (e.g. NGC 6337 and Sab 41) strongly suggest a binary origin for LIS. LIS seem confined to either the orbital plane as radially distributed knots or filaments, or to the polar regions as (mostly) low surface brightness jets triggered by a dynamo (Nordhaus \& Blackman 2006). A likely binary formation scenario includes the distribution of neutral clumps of dust and H$_2$ during the CE phase into the orbital plane that are then photo-ionised by winds during the PN phase. Deep [NII] imaging of the post-CE sample may reveal very low surface brightness jets to be more common than even found in this work (e.g. Vaytet et al. 2009). Evolutionary effects can probably explain the absence of LIS in some post-CE nebulae. 
   \item A preliminary investigation into a larger sample of nebulae with LIS found two thirds of them to have emission-line nuclei. A binary origin may therefore be responsible for LIS around emission-line nuclei whereby one or more CE phases created the identifiably post-CE morphology and dual-dust chemistry (De Marco \& Soker 2002; De Marco 2008). The most extreme LIS appear in nebulae surrounding [WO] nuclei distinguishing them apart from from all other PNe. Indirect evidence in support of these scenarios is provided by Hb 4 which exhibits jets and filamentary LIS, a [WO3] nucleus and dual-dust chemistry.
\end{itemize}

\begin{acknowledgements}
    BM acknowledges the support of an Australian Postgraduate Award and further support from Macquarie University, Strasbourg Observatory and PICS. 
    AFJM is grateful for financial assistance to NSERC (Canada) and FQRNT (Qu\'ebec). 
    We thank Anna Kovacevic for discussions, Bruce Balick for establishing the superb Planetary Nebula Image Catalogue used in this research, and Romano Corradi for kindly providing his FITS images of A 41. We acknowledge the conscientious and helpful support of St\'ephanie C\^ot\'e, Etienne Artigau, Claudia Winge, Simon O'Toole and Steve Margheim during the course of our Gemini programs.
    
    This research has made use of SAOImage \textsc{ds9}, developed by Smithsonian Astrophysical Observatory, the SIMBAD data base operated at CDS, Strasbourg, France, and made use of data products from the 2MASS survey, which is a joint project of the University of Massachusetts and the Infrared Processing and Analysis Centre/California Institute of Technology, funded by the National Aeronautical and Space Administration and the National Science Foundation. 
    Based on observations obtained at the Gemini Observatory, which is operated by the
    Association of Universities for Research in Astronomy, Inc., under a cooperative agreement
    with the NSF on behalf of the Gemini partnership: the National Science Foundation (United
    States), the Science and Technology Facilities Council (United Kingdom), the
    National Research Council (Canada), CONICYT (Chile), the Australian Research Council
    (Australia), Minist\'erio da Ci\^encia e Tecnologia (Brazil) and Ministerio de Ciencia, Tecnolog\'ia e Innovaci\'on Productiva (Argentina).
    This paper also made use of data obtained from the Isaac Newton Group Archive which is maintained as part of the CASU Astronomical Data Centre at the Institute of Astronomy, Cambridge. 

\end{acknowledgements}

\end{document}